\documentclass[12pt]{article}

\usepackage{epsfig}
\usepackage{color}
\usepackage{cite}
\usepackage{graphicx}
\usepackage{amssymb}
\usepackage{amsmath}
\usepackage{latexsym}

\newcommand{\be}{\begin{eqnarray}}
\newcommand{\ee}{\end{eqnarray}}
\newcommand{\im}{{\rm Im}}

\newcommand{\tr}{{\rm Tr}}

\newcommand{\half}{\frac{1}{2}}

\renewcommand{\d}{\delta}
\renewcommand{\dag}{\dagger}

\newcommand{\nn}{\nonumber}

\newcommand{\ep}{\epsilon}

\newcommand{\en}{\end{eqnarray}}

\renewcommand{\d}{\partial}

\newcommand{\bea}[1]{\left(\begin{array}{#1}}
\newcommand{\ena}{\end{array}\right)}

\def\norm@note#1#2{\special{}
  \ifinner{\ifdim\baselineskip=\z@
    \baselineskip18\p@\fi
    \ifhmode
    \raisebox{.5\baselineskip}[\z@][\z@]{%
      \rlap{\sf\scriptsize #2}}%
    \else\vskip-\baselineskip%
    \raisebox{-.6\baselineskip}[\z@][\z@]{%
      \rlap{\sf\scriptsize #2}}%
    \fi}%
  \else\marginpar{\raggedright\if@twoside\ifodd\c@page%
    \raggedleft\fi\fi\sf\scriptsize #1#2}%
  \fi}%

\begin{document}

\begin{titlepage}

\vskip.5cm
\begin{center}
{\huge \bf Sequestering  CP Violation and GIM-Violation  with Warped Extra
Dimensions} \vskip.1cm
\end{center}
\vskip0.2cm

\begin{center}
{\bf Clifford Cheung, A. Liam Fitzpatrick, Lisa Randall}
\end{center}
\vskip 8pt
\begin{center}
{\it Jefferson Physical Laboratory \\
Harvard University, Cambridge, MA 02138, USA} \\
\vspace*{0.3cm}
{\tt cwcheung@fas.harvard.edu, fitzpatrick@physics.harvard.edu,
randall@physics.harvard.edu}
\end{center}

\vglue 0.3truecm

\begin{abstract}
\vskip 3pt
\noindent
   We propose a model of spontaneous CP violation to address the
strong CP problem in warped extra dimensions that relies on
sequestering flavor and CP violation. We assume that
brane-localized Higgs Yukawa interactions respect a $U(3)$ flavor
symmetry that is broken only by bulk fermion mass  and Yukawa terms.  All CP
violation arises from the vev of a CP-odd scalar field localized in
the bulk. To suppress radiative corrections to $\bar{\theta}$, the
doublet quarks in this model are localized on the IR brane.  We
calculate constraints from flavor-changing neutral currents (FCNCs),
precision electroweak measurements, CKM unitarity, and the electric
dipole moments in this model and predict $\bar{\theta}$ to be at 
least about  $10^{-12}$.

\end{abstract}

\end{titlepage}

\newpage

\section{Introduction}
\label{sec:intro}

\definecolor{MyDarkRed}{rgb}{0.45,0.0,0.05}
CP violation in the kaon system was first observed more than 40
years ago \cite{Christenson:1964fg}, and yet no evidence of CP violation in the
strong interactions has since been observed.
Whereas the weak phase is of order unity, the strong CP angle
$\bar{\theta}$ is constrained to be $\lesssim 10^{-10}$.

 The quarks get their masses in the Standard Model 
 from two Yukawa matrices $Y_u, Y_d$
for the up-type  quarks and down-type quarks, respectively,
which induce a unitary CKM matrix when rotated into
the mass eigenbasis:
\be
\lambda_u &\rightarrow& L_u \lambda_u R_u^\dagger = \mu_u / v \nn\\
\lambda_d &\rightarrow & L_d \lambda_d R_d^\dagger = \mu_d / v \nn\\
V_{\rm CKM} &=& L_u L_d^\dagger \label{eq:basischange} \ee $V_{\rm
CKM}$ contains a single phase that cannot be removed by vector-like
rotations of the quarks. This phase accounts for all observed CP
violation in the Standard Model.  
  The rotation in equation \ref{eq:basischange} is generically chiral
and thus shifts $\theta_{\rm QCD}$, but in such a way that the
physically observable $\bar{\theta}$ remains invariant \be
\bar{\theta} &=& \theta_{\rm QCD} - \arg \det \lambda_u \lambda_d
\label{eq:thetabar} \ee The renormalizable operator
$\theta_{\rm QCD} G \tilde{G}$ could in principle violate CP
independently of the weak interactions. Without some mechanism for preventing strong CP violation, the cancellation
between $\theta_{\rm QCD}$ and ${\rm arg det} \lambda_u \lambda_d$
appears extremely fine-tuned.

An additional difficulty with constraints on CP
violation  occurs in models where the flavor structure of the Standard Model
is explained by UV physics. Since flavor and CP
in the Standard Model have the same origin (the Yukawa matrices),
any solution to the flavor problem is likely to also introduce new sources of
CP violation (CPV). Integrating out heavy fields generically induces
dimension-5 operators that contribute to the neutron electric dipole
moment (EDM), independently of $\bar{\theta}$.  Axion models, which address 
only $\bar{\theta}$ have
nothing to say about these additional contributions. Of course the order of magnitude involved is far less severe, but nonetheless these additional contributions to EDMs generally require additional model-building constraints.




The approach we take addresses these issues and falls in the class
of solutions referred to as spontaneous CP violation. The idea is
that CP is a valid symmetry in the UV, but is broken spontaneously
in such a way that phases enter $V_{\rm CKM}$ but not $\arg \det
\lambda_u \lambda_d$.


In this paper we consider a theory of spontaneous
CPV embedded in RS.
We utilize in a warped scenario
the extra dimensional mechanism of \cite{schwartz},
which ``sequesters'' the source of CPV from operators that could
directly transmit it to $\bar{\theta}$.
Thus, we are able to address the strong CP problem even with
new KK modes near the electroweak scale, though electroweak
constraints push the KK masses to be $\gtrsim 15-18$ TeV.
This is low from the point of view of models addressing
the strong CP problem, though unfortunately it is fairly
high in terms of addressing the gauge hierarchy
problem.  
 However we show that 
embedding a solution to the strong CP
problem based on spontaneous CPV inside RS allows us to explain mass scales and further suppress some potentially dangerous CP and flavor violating contributions, thereby allowing us to address both   the gauge hierarchy and fermion mass hierarchy
problems.

   Furthermore, sequestering CPV  relaxes constraints on the KK scale of RS from
neutron electric dipole moment (EDM) measurements.
We will show that if CPV is sourced in the bulk and transmitted
through the fermion wavefunctions, then the flavor violation and
CP violation  very efficiently
``washes out'' of the excited modes but not the
zero modes.  This is to be contrasted with
``anarchic'' models of flavor in RS, where the CPV comes from
Higgs Yukawas with ${\cal O}(1)$ entries and phases, and thus is equally
present at every KK level.  We show that our model suppresses
EDM operators radiatively generated from integrating out KK modes.

As we will explain in more detail throughout the paper, the suppression of flavor and CP violation in our model exploits four ingredients
that are natural in the context of extra dimensions:

\begin{enumerate}

\item \emph{AdS spacetime isometries in the bulk.}
The $SO(4,2)$ symmetry of AdS greatly restricts the possible couplings
of  bulk fermions a bulk scalar field $\Phi$, analogous to the restrictions
imposed by 5D Lorentz symmetry in flat-space \cite{schwartz}. 
We assume CP is spontaneously broken when $\Phi$ gets a vev. This generates a
\emph{hermitian} bulk mass matrix $\mathcal{M}$ for the fermions as with the mechanism of \cite{schwartz}. 
 The zero mode
wavefunctions $F^{(0)}(\phi)$
at a position $\phi$ in the bulk then take the form
\be F^{(0)} \propto \mathcal{P} \exp \left(\int^\phi
   \mathcal{M}(\phi')
d\phi' \right), \ee
where $\mathcal{P}$ denotes path ordering with respect to $\phi$. This
has real determinant since $\mathcal{M}$ is Hermitian.
Taking $\lambda_{u,d}$ to be the brane Yukawas, which are real-valued
by CP,  one finds that $Y_{u,d} = F_q^{(0)} \lambda_{u,d} F_{u,d}^{(0)}$.
Both have real determinant, and $\bar{\theta}$ vanishes at tree-level.

\item  \emph{Sequestering of the CP violating scalar
field from  the boundaries and the local Yukawa interactions}.
Ingredient (1) enforces that CP violation in the bulk that is
transmitted through the fermion wavefunctions will not induce
$\bar{\theta}$ at tree level.  However, if the source of CPV has
non-negligible overlap with the TeV brane, then it can enter the
Yukawas through the direct coupling
 $\Phi_{ij} (\bar{Q}_L)_i H (U_R)_j$, which contributes order one
strong CP phases once $\Phi$ gets a vev.
 $\bar{\theta}$ would not be protected from these
contributions, and thus it is crucial that the source of CPV be
geographically sequestered from the Higgs Yukawas.  Furthermore, the
AdS isometries are explicitly broken at the boundaries of RS, and the coupling
of $\Phi$ to the fermions is not protected there. We therefore
assume that $\langle \Phi \rangle$ is localized somewhere in the
bulk, away from the boundary branes.  We will see that the farther
we localize $\Phi$ in the UV, the less its effect will be
on the KK fermion wavefunctions, so CP violation is even more effectively sequestered than one might naively assume.

For simplicity, we sequester $\Phi$ by
localizing it to a third brane in the warped bulk. It is possible to
imagine other equally viable profiles for $\Phi$.
It will be important that
the effect of $\Phi$ washes out much more efficiently
of the IR-localized KK mode wavefunctions
than in those of the  UV-localized zero modes so that there is 
a limit where the KK mode wavefunctions are
diagonal in flavor indices (up to negligible corrections) and
all CP violation enters the KK-reduced theory only through
the zero mode wavefunctions.

\item \emph{A large flavor symmetry. } At tree-level, the only aspect of the flavor
symmetry that is necessary for $\bar{\theta}$ suppression is that
all the up-type quarks get their masses at the same place in the
bulk, and similarly for the down-type quarks. This is generically a
feature of RSI models, where the Higgs vev must be localized near
the TeV brane in order to solve the hierarchy problem.
 To address loop corrections, we  assume that the flavor group
 $U(3)_Q \times U(3)_U \times U(3)_D$ breaks  to a
 diagonal $U(3)$
through the Higgs Yukawas, and further down to the $U(1)^3$ subgroup by
constant bulk fermion masses, and finally broken completely by a CPV
source in the bulk.  

We further assume that the bulk masses for the up-type singlets and down-type
singlets approximately commute, so that they are diagonal in approximately
the same basis\footnote{This could be achieved for instance by coupling together
the fields responsible for generating their bulk masses so that their
potential is minimized when they align.}.  In this case, the bulk masses
leave invariant an approximate $U(1)^3$  for each generation. It is not necessary to make any further assumption about
the bulk mass matrices aligning with the brane Yukawas.

\item \emph{Doublets confined to the IR brane.}
The doublets are confined to the
IR brane.  Consequently, the only states
charged under $SU(2)_L$ are zero modes.
This implies that
all fermion interactions with charged
currents are controlled by a single
CKM matrix $V_{\rm CKM}$ that is a generalization
of the Standard Model version.  
 In the Standard Model, the CKM matrix
is unitary since all of the left-handed fields
couple to the $W$ bosons with equal strength.
However, in 5d models, there are left-handed
fields, namely the left-handed KK excitations
of the electroweak singlets, which do not
couple to $W$ bosons.
Consequently, $V_{\rm CKM}$  is of the form
\be
V_{\rm CKM} &=& L_u P_0 L_d^\dagger
\en
where $P_0$ is a projection onto the
left-handed \emph{zero modes} and
$L_{u,d}$ are diagonalization
matrices that mix zero and KK modes.

With  the doublets to be confined to the IR brane, 
all CP and flavor violation appears through the singlet wavefunctions.
In particular, the singlet zero mode wavefunctions are responsible
for generating small Standard Model CKM angles.  When the CPV source
is located sufficiently far in the UV, the CKM angles appear generically as
fractional powers of ratios of fermion masses, and thus are generated 
naturally. 

\end{enumerate}

We will see that a consequence of these assumptions is that up to $O(v/M_{KK})^2$, we have an approximate GIM mechanism. 
Our basic outline for this paper is as follows. In section 2, we
review flavor in warped extra-dimensional models
and describe some differences in our model.
 We present our model and some of its features in section 3. In
sections 4 we discuss the KK reduced theory and its
interactions.  In section 5, we calculate the CP and flavor
constraints, and in section 6 we conclude.

\section{Flavor Physics in Higher-Dimensional Models}

Because the spontaneous breaking of CP is intimately connected with
the breaking of flavor symmetries in our model, we give a brief
aside about flavor physics in higher-dimensional models and RS in
particular.
The models we compare all have an extra dimension bounded by branes.
The original RS model had the entirety of the SM localized on the IR
brane. Confinement of the SM fields to the IR brane was assumed to
happen through some unspecified dynamics whose effects on flavor-changing,
etc., were impossible to calculate. Putting the SM gauge fields in the
bulk (first considered in \cite{Pomarol:1999ad,Davoudiasl:1999tf}) allowed the
possibility for gauge coupling unification\cite{Randall:2001gb,Randall:2001gc,
Goldberger:2002hb,Agashe:2005vg,Agashe:2002pr},  but it had calculable corrections to
precision electroweak observables that raised the scale of KK gauge bosons to
23 TeV.  Putting all the fermions in the bulk drastically weakened such constraints,
and further allowed for the warping of fermion wavefunctions to explain the hierarchy
in their masses \cite{Chang:1999nh,neubert,agasheperez,huber} while the brane
 Yukawas
and masses were order ${\cal O}(1)$ and anarchic. 
 In these ``anarchic'' models,
the small zero mode masses and CKM angles arise from the hierarchy in 
fermion wavefunctions on the TeV brane, as follows.  The Yukawa matrices in
the KK reduced theory are $Y_{u,d} = F_{Q}^{(0)} \lambda_{u,d} F_{u,d}^{(0)}$, 
where $\lambda_{u,d}$ are the 5d brane Yukawa matrices with ${\cal O}(1)$ values and
phases in every entry and $F_{Q,u,d}^{(0)}$ are matrices of the zero mode
wavefunctions at the IR brane.  The zero mode masses then are dynamically generated
as 
\be
m_{u,d} &\approx& v F_Q^{(0)} F_{u,d}^{(0)} 
= v \sqrt{\frac{1+2\nu_Q}{1 - e^{-(1+2\nu_Q)k\pi r_c}}}
 \sqrt{\frac{1+2\nu_{u,d}}{1 - e^{-(1+2\nu_{u,d})k\pi r_c}}}
\en
where $\nu_{Q,u,d}$ are the bulk masses in units of the AdS curvature scale $k$,
and we have used the zero mode wavefunctions $F^{(0)} = 
\sqrt{(1+2\nu)/(1-e^{-(1+2\nu)k\pi r_c})}$.
Therefore, a modest spread in the values of $\nu$ for the different fermions
gives rise to an exponential hierarchy in the masses.  Furthermore, the 
CKM angles are controlled by the bulk masses for the doublets, as follows. 
The CKM matrix diagonalizes $Y_{u,d} Y_{u,d}^\dagger =
F_{Q}^{(0)} \lambda_{u,d} F_{u,d}^{(0)2} \lambda_{u,d}^\dagger
F_{Q}^{(0)}$, and thus the CKM angles are approximately
\be
\theta_{ij} &\approx& \frac{F_{Qi}}{F_{Qj}}
\en

 Constraints on
the $T$ parameter from bulk SU(2) gauge boson contributions were further reduced
in models in which
global custodial $SU(2)$ was extended to a gauge symmetry in the bulk.
Potential problems with RS models of flavor
 are the $Z$ to $b \bar{b}$ coupling, which is modified
because $b_L$ cannot be localized too far away from the IR brane,
where KK modes of bulk gauge fields are concentrated, as well as
large $\epsilon_K$
 and the electric dipole moment of the neutron.  These latter constraints are
particularly bad because of the parity symmetry in the bulk, which
means that both left and right-handed fields couple to the
$W$ gauge boson KK modes. These constraints are quite severe\cite{agashecontino,
agashesundrum}.

\begin{figure}[t]
\begin{center}
\includegraphics[width=0.47\textwidth]{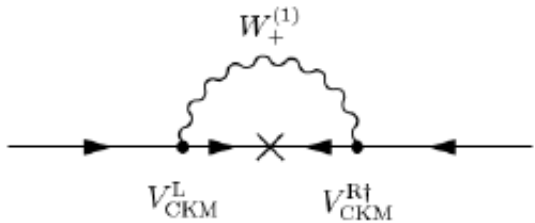}
\end{center}
\caption{
A one-loop contribution to $\bar{\theta}$ in the case where
both the doublets and the singlets are in the bulk.  We therefore
take the doublets to be localized on the brane.}
\label{fig:thetabaroneloop}
\end{figure}

As in any model of flavor, one can also try to address the strong CP
problem in these models.
The first proposal to use twisting to solve the strong CP problem
was made in \cite{schwartz}, in a flat extra dimension.  There, CP
violation occurred because up and down type quarks were located at
opposite ends of the fifth dimension so that CPV couldn't be
eliminated from both simultaneously. However, RS requires that all
Yukawas be on the IR brane, so it is necessary to generate
different phases for the up and down type right handed fields,
which can be done by placing them both in the bulk rather than on
branes. Because of the strong EDM constraint, we do not allow left
handed Standard Model fermions in the bulk but instead sequester
them on the IR brane. The reason is that bulk doublets have a
vector-like tower of KK modes, so that there exist right-handed
fields charged under $SU(2)_L$.  Then, there is not only a CKM
matrix for the left-handed fields, but an additional CKM matrix
for the right-handed fields as well,
 and $\bar{\theta}$ would be renormalized at
one-loop by the diagram in figure \ref{fig:thetabaroneloop}. We thus consider
brane-localized doublet fermions only.

However, as mentioned,  models with light fermions localized on the
TeV brane face very restrictive constraints from precision electroweak
measurements, since the KK gluon wavefunctions are enhanced by a volume
factor in the IR. Our strongest constraints will be from these
measurements, which puts the scale of KK gauge bosons in our
model at about 15-18 TeV.
Similar though weaker constraints come from flavor and unitarity
of the observed CKM matrix for the zero mode quarks.

It is nonetheless worth noting the interesting flavor physics of this model
 with bulk right handed fermions but IR-localized 
left-handed fermions. We will assume a $U(1)^3$
 symmetry is preserved by the bulk masses which is broken only
by a localized  CPV field in the bulk coupling to the bulk singlets
through Yukawa couplings. Unlike in ``anarchic'' models, 
the brane Yukawa matrices in our model are flavor universal and thus
the CKM matrix arises fairly naturally
from the different values of the right-handed fermion wave
functions, even though their bulk masses are similar and the vev
of the CPV field is quite generic. 
More precisely, we
will show that in our setup the Yukawas are given by
\be
Y_{u,d} &\approx&
     F_{u,d}^{(0)} = \left( \frac{ 1+2\nu_{u,d}}{1-e^{(1+2\nu_{u,d})k\pi r_c}}
\right)^{ 1- { \phi_0 \over \pi} } e^{k r_c g_{u,d} \Phi}  
\left( \frac{ 1+2\nu_{u,d}}{1-e^{(1+2\nu_{u,d})k\pi r_c}}
\right)^{  \phi_0 \over \pi} 
\en
where $\phi_0$ is the location of the CPV source $\Phi$, with the
usual convention that $\phi=0$ ($\pi$) is the UV (IR) brane, and
$g_{u,d}$ is the coupling strength of $\Phi$ to the bulk fermions. 
For ${\cal O}(1)$ values of $k r_c g_{u,d} \Phi$, 
this form of the Yukawas  generates CKM angles that come dominantly from
the down-type Yukawa and are approximately
\be
\theta_{ij} &\approx& \left( \frac{m_{di}}{m_{dj}} \right)^{1- {\phi_0 \over \pi}},
\en
which works rather well for reproducing the observed values.  

Another nice feature
of RS flavor models in general is that the KK mode mass matrix is
close to the identity. This helps with CP violating and flavor
violating effects. Independently of CP, the flavor model might be worth
pursuing,
 particularly if the stringent $Z$ boson
coupling constraints can be alleviated.

Finally, we compare the role of twisting in the  model of
\cite{schwartz} and our model. Twisting plays an essential role in
the first model since both the up and down mass matrices can be
diagonalized by different transformations at different points and it
is only because of the different phase rotations required at the
different points in the bulk that CP violation occurs.  The use of
twisting is a little more subtle in our case, where
it generates a hierarchy in the CKM angles and furthermore sequesters C
violation from the IR brane Yukawa couplings which is essential to
suppressing strong CP violation.

\section{The Model}
\label{sec:model}

\subsection{Definition of the Model}

RS is defined on a non-factorizable warped geometry with metric
\be ds^2 &=& e^{-2kr |\phi|} \eta_{\mu\nu}dx^\mu dx^\nu + r^2
d\phi^2, \ee where the extra dimension is an $S^1/Z_2$ orbifold of
size $r$ labeled by a coordinate $\phi\in [-\pi,\pi]$,such that
the points $(x^\mu, \phi)$ and $(x^\mu,-\phi)$ are identified.  At
orbifold fixed points at $\phi=0$ and $\phi=\pi$ lie the UV and
IR brane, respectively. Note that this orbifold boundary condition
forces either the left or right-handed zero mode of every bulk
fermion wavefunction to vanish.

The singlet SM fermions, $(U_L',U_R)_i$ and $(D_L',D_R)_i$,
come in Dirac multiplets that live in the bulk, while
the doublets $(Q_L)_i$, are localized on the
brane.  Here $i$ is a
flavor index and $'$ labels the fermion component whose zero mode
is projected out by the orbifold boundary condition. Thus, zero
modes exist only for $Q_L$, $U_R$, and $D_R$, and not the primed fields.
The Higgs, $H$, is
localized on the IR brane. Finally, we include a CP odd bulk
scalar, $\Phi_{ij}$, that is localized at a position $\phi_0$ in the
bulk and charged under flavor and acquires a
complex vev that is the source of all CP phases in the theory. The
action for our theory (neglecting kinetic terms) is
\be \label{eq:OurAction} S &=& S_{\rm bulk} + S_{\rm brane},\\
S_{\rm bulk} &=& \int d^5 x \sqrt{G} \left\{ -\bar{U}_L'(M_u + g_u \Phi
\delta(\phi-\phi_0)) U_R \right. \\
&& \qquad\qquad\quad \left. -\bar{D}_L'(M_d + g_d \Phi
\delta(\phi-\phi_0)) D_R +{\rm h.c.}\right\},\\
S_{\rm brane} &=& \int d^4 x \sqrt{-g} \left\{ \bar{Q}_L \tilde{H}
\lambda_u  U_R + \bar{Q}_L H  \lambda_d  D_R + {\rm h.c.} \right\}
, \ee where $g_{u,d}, \lambda_{u,d}$ are numbers and $M_{u,d}$ are
matrices. Note that we have assumed
that CP is a valid symmetry in the UV, so \emph{all} of the
couplings in this action are manifestly real.

The mechanism for eliminating $\bar{\theta}$ at
tree-level relies crucially on the spacetime symmetry, $SO(4,2)$,
which constrains the only renormalizable interaction between
the CPV source $\Phi$ and a bulk fermion $\Psi$ to be of the form
 $\Phi_{ij} (\bar{\Psi}_{Li} \Psi_{Rj} + \bar{\Psi}_{Ri} \Psi_{Lj})$,
so only the hermitian piece of $\Phi$ couples to the fermions in the
action.
If $\Phi$ is
furthermore $CP$ odd, as we impose,
then its
couplings are further restricted: \be {\cal L} &\supset& g(\Phi_{[ij]} +
\Phi_{[ij]}^\dagger) (\bar{\Psi}_{Li} \Psi_{Rj} + \bar{\Psi}_{Ri}
\Psi_{Lj} )\delta(\phi-\phi_0) \en
After $\Phi$
acquires a vev, this will introduce phases and twisting in the
fermion wavefunctions.

\subsection{Zero Mode Wavefunctions}

In this section we show that zero modes wavefunctions generically
have real determinant in extra dimensional theories and establish
our notation for the fermion wavefunctions that we will use
in the remainder of the paper.  For now, let
us consider just a single massive bulk fermion flavor multiplet
$\Psi_i$ coupled to a CP-odd bulk scalar $\Phi$; of course, our
results will hold for the SM fermions, where $\Psi_R = U_R,D_R$ for
the up-like and down-like singlets, respectively. The action is
\be S &=& \int d^5x \sqrt{G} \left\{ \frac{1}{2} i \bar{\Psi}_i
\Gamma^A(\partial_A - \overleftarrow{\partial}_A) \Psi_i -
\bar{\Psi}_i (M_{ij}+g \Phi_{ij}\delta(\phi-\phi_0))\Psi_j \right\}. \ee
After imposing orbifold boundary conditions on the bulk fermions,
the KK reduction will contain a vector-like spectrum of KK
excitations for all but a single chiral zero mode. As before,
\emph{all} the couplings in the above action are real because CP
is preserved in the UV.  We assume that $\Phi$ has some CP
respecting potential, $V(\Phi)$, which is therefore even
in $\Phi$,
that induces spontaneous symmetry breaking, giving $\Phi$ a
complex vev that sources CP violation in our theory.

We perform a standard KK reduction of the bulk fermions as follows:
\be \Psi_{L,R}(x,\phi)_i &=& \frac{e^{2kr|\phi|}}{\sqrt{r}}
\sum_\alpha F_{L,R}(\phi)_{i\alpha} \times
\psi_{L,R}(x)_{\alpha}.
\label{eq:kkreduction}\ee

The explicit factor of $e^{2kr |\phi|}$ is conventional, and
simplifies the expressions for the normalization conditions of the
wavefunctions. Let us consider the KK reduction of $\Psi$. All of
the $x$ dependence lies in the (vector of) dynamical 4D fields
$\psi$, while the $\phi$ dependence is included in the (matrix of)
wavefunctions $F$. Notice that this matrix has a flavor index $i$
which runs from 1 to 3 as well as a KK mode index $\alpha$, which
runs from 1 to $N$, where $N$ labels the KK mode at which the model
becomes gravitationally strongly coupled. In general, it is very
natural to group these KK modes in groups of threes because of the
approximate flavor symmetry. For example there will always be three
exactly massless zero modes (before electroweak symmetry breaking),
which in the case of $\Psi = U_R$ correspond to the right-handed up,
charm, and top.  Thus, we can parse the mass matrix into a separate
3 by 3 matrix for each KK level: \be F_{i\alpha} &=&
  \left\{ \overbrace{ \{ F_{i1}, F_{i2}, F_{i3} \}  }^{\textrm{zero  modes} },
   \overbrace{ \{ F_{i4}, F_{i5}, F_{i6} \} }^{ \textrm{ 1st  excited  modes} },
\dots \right\} \\
  &=& \left\{ F_{i\alpha}^{(0)}, F_{i\alpha}^{(1)}, \dots
  \right\}.
\ee Using this notation, it is straightforward to refer to a
particular KK mode of a particular generation: for example
$F_{i3}^{(1)}$ denotes the 3-vector wavefunction corresponding to
the first KK excitation of a third generation fermion. \footnote{
There is some freedom here to choose what we mean by ``third
generation'' of the excited modes, since the mass matrix remains
diagonal under permutations of any two fermion modes. One sensible
choice would be to rank the ``generations'' by increasing mass
within each KK level.  Another reasonable definition would be to
demand that as $\Phi$ is decreased continuously to zero, the $F$
matrices transform continuously to diagonal matrices. We choose the
latter of these, though clearly none of our results depend on this
choice.}   Also, note that in a theory in which all the flavors are
decoupled from each other, $F^{(n)}$ is simply a diagonal matrix. In
a theory in which the flavors are degenerate, $F^{(n)}$ is
proportional to the identity matrix.

For each KK fermion $(\psi_{L,R})_\alpha$, its wavefunction
$(F_{L,R})_{i\alpha}$ obeys the first order wave equation
\be \left(\pm \frac{1}{r}\partial_\phi - (M+g \langle \Phi \rangle
\delta(\phi-\phi_0)
)\right)_{ij}(F_{L,R})_{j\alpha} &=& - e^{kr|\phi|}  m_\alpha
(F_{R,L})_{i\alpha} 
\label{eq:fullwaveequation}\ee
where $m_\alpha$ is the KK mass.
Going to a more convenient variable \cite{neubert}, $t=\epsilon
e^{kr|\phi|} \in [\epsilon,1]$, where $\epsilon = e^{-kr \pi} =
10^{-16}$, rescaling $F_{L,R}(\phi) \rightarrow \sqrt{k r
\epsilon} F_{L,R}(t)$, and defining
\be \label{eq:nudef} \mu_{ij} &=& \frac{1}{k}\left(M + g kr t
\langle \Phi \rangle \delta(t-t_0)\right)_{ij} , \\
x_\alpha &=& \frac{m_\alpha}{\epsilon k}. \ee
 Now because
of the orbifold boundary condition, we are forced to set either
the zero mode for $\Psi_L$ or $\Psi_R$ to vanish at the orbifold
fixed points $t = \epsilon,1$.  Assuming we choose to eliminate
the right-handed zero mode, the left-handed zero mode then
satisfies the wave equation and orthonormality condition
\be \left(\partial_t - \frac{\mu}{t}\right)F_L^{(0)}&=&0 
\label{eq:zeromodewaveequation}\\
\int_\epsilon^1 dt F^{(0)\dagger}_L F^{(0)}_L =
   \int_\epsilon^1 dt F^{(0)\dagger}_R F^{(0)}_R &=&
  {\bf 1}, \ee
where we have suppressed flavor indices.  Equation 
\ref{eq:zeromodewaveequation} can be
formally integrated to obtain
\be F_L^{(0)} \propto \mathcal{P} \exp \left(\int_\epsilon^t \frac{\mu(t')}{t'}
dt' \right),
\label{eq:zeromodeint}\ee
where $\mathcal{P}$ denotes path ordering with respect to t.
The proportionality constant in \ref{eq:zeromodeint} is a constant
hermitian matrix.   We have chosen the lower bound on the integral
 in \ref{eq:zeromodeint} to be $\epsilon$ so that the normalization
matrix is close to the identity for UV localized zero modes, and 
thus does not have a qualitative effect on any of our discussions.
We do of course include it in all of our numeric computations.
Because $\mu$ is Hermitian, we see that $F_L^{(0)}$ is a product
of infinitesimal Hermitian matrices, so it has real determinant.
Thus, extra dimensions generically yield zero mode matrix
wavefunctions which have real determinant.

\subsection{Approximate Dependence of CKM angles on $t_0$}

Consider now the structure of the observed Yukawa
matrices if
 we assume that $\Phi$ has sufficiently
large entries to completely scramble the
entries of any matrix that it multiplies.
Let us split the path-ordered integral in equation
\ref{eq:zeromodeint} into three pieces:
\be
F^{(0)}(1) &=& F(1;t_0)
   \times \exp \left( kr g \Phi \right)
 \times
   F(t_0;\epsilon) \nn\\
  F(1;t_0) &\equiv& \left[ \mathcal{P} \exp \left(
   \int_{t_0}^1 \frac{\mu}{t'} dt' \right) \right] = t_0^{-\nu}\nn\\
  F(t_0;\epsilon) &\equiv&
  \left[ \mathcal{P} \exp \left(
   \int_{\epsilon}^{t_0} \frac{\mu}{t'} dt' \right) \right]= 
\left(\frac{\epsilon}{t_0} \right)^{-\nu} 
\en
where $\nu = M/k$ is a constant matrix.  Then,  $L_{u,d}$
restricted to the zero modes approximately
diagonalizes
\be
Y_{u,d} Y_{u,d}^\dagger &=&
F_{u,d}(1;t_0) e^{krg \Phi} F_{u,d}(t_0;\epsilon)
F_{u,d}(t_0;\epsilon)^\dagger
 e^{krg\Phi} F_{u,d}(1;t_0)^\dagger
\en
Assume for simplicity that the nonzero entries of $\Phi$
are not so large that they contribute to the hierarchy
of the fermion wavefunctions on the TeV brane;
one could of course consider more general cases.
Since $e^{krg \Phi}$ completely mixes flavor,
the above expression has the basic structure
$(Y_{u,d} Y_{u,d}^\dagger)_{ij} \approx
F_{u,d}(1;t_0)_{ik} \zeta_{kk'}  F_{u,d}(1;t_0)_{k'j}^\dagger$ with
$\zeta_{kk'} \sim {\cal O}(1)$, which
leads to mixing angles of size $\theta_{ij} \approx
(m_i/m_j)^{-(\log t_0)/(k\pi r_c)}$ since $F(1;t_0)$ contains
only a fraction $\log t_0/\log \epsilon$ of the entire
fermion mass hierarchy.  Because the down-type masses are
less hierarchical, they will give larger contributions
to the CKM angles than the up-type masses will.  Then, we
are left to ask whether there is a value for $t_0$ such
that $(m_d/m_s)^{-(\log t_0)/(k\pi r_c)} \sim \lambda_C$ and
$(m_s/m_b)^{-(\log t_0)/(k \pi r_c)} \sim \lambda_C^2$,
where $\lambda_C = |V_{us}| = 0.23$.  For instance, running
the quark masses up to the scale $Q=10 $ TeV allows
$m_d = 3.5 MeV, m_s = 30 MeV$, and $m_b = 2.2$ GeV
\cite{huber} within experimental uncertainty, and these
values satisfy $(m_d/m_s)^2 \approx (m_s/m_b)$. The further
condition $(m_d/m_s)^{-(\log t_0)/(k\pi r_c)} = 0.23$ implies
$t_0 \approx 4 \times 10^{-11}$.
Such a value puts the CPV source very far in the UV and therefore
 will have a very small effect
on the excited mode wavefunctions.
Of course deviations from complete genericity are possible but
as we will see unnecessary.  

\subsection{Specific Values}
\label{sec:values}
For concreteness, 
we present here  a specific choice of $\Phi$ and bulk
masses that roughly reproduces the Standard Model fermions masses
and KM matrix. \be
\nu_U &=& (-0.831, -0.665, -0.241) \nn\\
\nu_D &=& (-0.788, -0.734, -0.632) \nn\\
\langle k r \Phi\rangle &=& \bea{ccc} 0 & 1.039 i & -1.342 i \\
   -1.039i & 0 & 1.481i \\
   1.342 i & -1.481 i & 0 \ena \nn\\
g_u &=& 0.3 \nn\\
g_d &=& 0.7 \nn\\
t_0 &=& 10^{-12}
\en
These numbers are not intended to be
an indication of what parameters are most expected
given some prior distribution. Rather, we want
to indicate what is possible with ${\cal O}(1)$ UV parameters.
We set the KK scale $\mu_{\rm TeV} \equiv \epsilon k$ to be 15$v$, where $v$ is
the Higgs vev 246 GeV, so that the KK gauge bosons are
at $\sim 9$ TeV, in order to demonstrate
that $\bar{\theta}$ is sufficiently small in this case.  As we discuss
in section \ref{sec:constraints}, electroweak constraints
force $\mu_{\rm TeV}$ to be around 20$v$, so that $\bar{\theta}$ does
not provide the strongest constraint on the model.

We obtain the following quark masses for the
zero mode and first excited modes:
\be
m_u &=& \textrm{( 1.6  MeV, 420  MeV, 176 GeV, 9.8 TeV, 10.3 TeV, 10.7 TeV)}
   \nn\\
m_d &=& \textrm{( 3.5 MeV , 41 MeV, 2.1 GeV, 9.6 TeV, 10.1 TeV,
   10.4  TeV)}
\ee
The absolute value of the KM matrix entries are
\be
 \bea{ccc} |V_{ud}| & |V_{us}| & |V_{ub}| \\
   |V_{cd}| & |V_{cs}| & |V_{cb}| \\
   |V_{td}| & |V_{ts}| & |V_{tb}| \ena &=&
  \bea{ccc} 0.982 & 0.185 & 0.0091 \\
   0.185 & 0.980 & 0.046 \\
   0.012 & 0.045 & 0.998 \ena
\en
and the rephasing invariant $\Delta^{(4)} \equiv
{\rm Im} ( V_{11} V^\dagger_{12} V_{22} V^\dagger_{21}) =
  7.6 \times 10^{-5}$   Because of mixing
with the excited modes, the CKM matrix for the zero modes
is not exactly unitary.  The first and second rows satisfy
\be
|V_{ud}|^2 + |V_{us}|^2 + |V_{ub}|^2 -1 &=& - 0.0021 \nn\\
|V_{cd}|^2 + |V_{cs}|^2 + |V_{cb}|^2 -1 &=& -0.0024
\en
which is at the limit of current bounds on unitarity \cite{ceccucci}
\footnote{See their section 11.4} and decreases proportional to 
$1/m_{\rm KK}^2$. This is further enhanced
by $\sum_{n=1}^\infty n^{-2} =\pi^2/6$ when we include several higher
KK modes.

\subsection{Fermion Wavefunction Dependence on the Bulk Masses}

The dependence of the KK fermion wavefunctions on their bulk
masses will be important for all of our phenomenological
constraints. In particular, as we will now discuss,
the value of the KK fermion wavefunctions in the IR
is very insensitive to flavor violation from
sources in the bulk, and especially in the UV.

Most of the qualitative features of the twisted excited mode
wavefunctions can be understood in the absence of twisting.  The
method we use for solving the wavefunctions with and without
twisting is reviewed in appendix \ref{sec:kkmodewf}. To begin, we
note that in the single-generation case, the left-handed mode
wavefunctions are given approximately by \be f_L(t) &=& \left\{
\begin{array}{cc}
  \frac{ \sqrt{2t} J_{-\half + \nu}(x t)}
{ J_{-\half + \nu}(x) }, & \nu > -\half \nn\\
  - \frac{ \sqrt{2t} J_{\half - \nu}(x t)}
{ J_{\half - \nu}(x) }, & \nu < -\half
\end{array} \right.
\en
except for the zero mode, for which $f_R$ vanishes and it is easy to
solve for $f_L$:
\be
f_L^0 &=& \sqrt{\frac{1 + 2\nu}{1 - \ep^{1+2\nu}}}
t^{\nu}
\en

An important point is that the wavefunctions of the excited modes
depend weakly on the values of the $\nu$'s, which are the source
of flavor-breaking.  Indeed, up to ${\cal O}(\epsilon)$ corrections,
the KK modes at the TeV brane are all
$\sqrt{2}$, independent of the value of $\nu$.  This
plays an important role in suppressing flavor-violation,
since it reduces the flavor-violation in the KK mode
wavefunctions.

The reason for this is that the behaviour near the UV boundary
approximately fixes the phase of the $f_L$ modes, so that
there is no $\nu$-dependent phase shift, as there
would be if we had constant bulk masses $\nu$ in flat space.
This follows from
the fact the $f_L$ are a linear combination of a regular mode
and a singular mode in the $t \sim \epsilon$ region, and the
regular mode vanishes at $t\rightarrow 0$.  The singular mode
is therefore very suppressed, and the turnaround point
(where $f'_L(t) = 0$) from singular to regular mode occurs
at very small $t$:
\be
t_{\textrm{turnaround}} &=&
   \left( \frac{4 \nu^2 - 1}{x^2} \epsilon^{-2\nu -1} \right)^{\frac{1}
{1-2\nu}} \qquad (\nu < -\half) \en so in the warped space case, the
phase shift due to $\nu$ is small and suppressed by powers of
$\epsilon$. Another difference is that there is less variation
required among the $\nu$'s in warped space, because of an
enhancement from $kr_c \sim 10$. The zero modes at the IR brane in
warped space are \be
f^{(0)}(t=1) &=& \sqrt{\frac{ 2\nu+1}{ 1 - e^{-(2 \nu+1) kr_c \pi}}} \\
   &\approx& \sqrt{ | 2 \nu + 1|} e^{-|\nu + \half |kr_c \pi} \qquad
  (\nu < -\half)
\en
and therefore we need $\nu + \half$'s of size
$\frac{1}{kr_c \pi} \log (( m_t/m_u)^{1/2}) = 0.17$.  This
might perhaps be considered  a little more tuned than in flat space,
since the $\nu$'s need to fall close to $-\half$ in warped space.
However, once the $\nu$'s are fixed from experimental constraints,
the KK mode wavefunctions are more flavor universal which leads
to much weaker phenomenological constraints.

\section{Interactions in the KK Reduced Theory}
\label{sec:vertices}

\subsection{Yukawa Interactions}
\label{sec:yukinteraction}

In this subsection we determine the flavor structure of the Yukawa
interactions.  In particular, we will show that just like the
charged $W$ interactions, the Yukawas can be written entirely in
terms of the CKM matrix.

To begin, we plug back in to Eq.~(\ref{eq:OurAction}), and find
that the effective 4D Yukawa interactions between zero modes
become \footnote{Here we denote fields in the KK reduced description
by lower-case letters, as in equation \ref{eq:kkreduction}.
For example, $u_R^{(i)}$ is in the KK tower of the 5D field
$U_R$, while $H$ and $h$ are the same since $H$ is a brane-localized
field in the 5D description.}
\be S_{\textrm{4D Yukawa}} &=& \int dx^4 \sqrt{-g}\left\{
\bar{q}_L \tilde{h} Y^{(0)}_u  u^{(0)}_R + \bar{q}_L h Y^{(0)}_d
d^{(0)}_R \right\}, \ee
where $Y^{(0)}_u$ and $Y^{(0)}_d$ are defined by
\be \label{eq:Ydefs} Y^{(0)}_u &=& \lambda_u F_u^{(0)} |_{\phi=\pi}, \\
Y^{(0)}_d &=& \lambda_d F_d^{(0)} |_{\phi=\pi}. \ee
and $\lambda_{u,d}$ are just real numbers.
As we showed explicitly in the previous section, in extra
dimensional scenarios, zero mode wavefunctions $F^{(0)}_{u,d}$
are complex but have real determinant.  Thus,
from
Eq.~(\ref{eq:thetabar}) we see that $\bar{\theta}$ vanishes at tree-level
while CP violation in the weak interactions does not.
In order to calculate the radiative corrections to this tree-level
result, we need to work out the interactions of the effective theory.
The KK tower of fermions has interactions that are simple
extensions of those of the Standard Model, though with some
non-trivial consequences. In the SM, the only source of masses is
the Yukawa interactions, and so going to the mass eigenbasis
pushes all CP violation into a single unitary KM matrix in the
gauge kinetic terms.  In contrast, extra dimensional models
contain Yukawa interactions that mix different KK levels, as well
as  KK mass eigenvalues  from the dimensional reduction.

Let us consider the effective 4D mass matrix, limiting the
following discussion to the zero modes and first excited modes.
Our arguments can be extended easily to the entire KK tower. We
will group the left and right-handed up-like modes together
\be u_L = (u_L^{(0)}, u_L^{'(1)}), \\ u_R = (
u_R^{(0)},u_R^{(1)}), \label{eq:nicebasis}\ee
where each is a vector of six fields: three for the zero mode,
and three for first excited modes. The ``zero'' mode for $u_L$ is
just the brane-localized up-type component of the doublet $q_L$,
whereas $u_R^{(0)}, u_L'^{(1)},u_R^{(1)}$ are zero and
KK modes of the bulk field
$U_R$
(recall that there are no bulk $q_L$'s).
Note that the ordering has been
chosen so that the mass matrix is approximately diagonal in this
basis.  The effective 4D mass matrix receives contributions from
Yukawa interactions and KK masses:
\be S_{\textrm{4D mass}}
&\supset& \int d^4 x \left\{
\bar{u}_L M_u u_R \right\}, \\
M_u &=& v Y_u+m_u,\\
 Y_u &=& \bea{cc}
  F_u^{(0)}   & F_u^{(1)}  \\
 0  & 0 \ena,\\
 m_u &=&\bea{cc}
  0 & 0   \\
 0 & m_u^{(1)} \ena
\label{eq:mudef}
\ee
where $m_u$ is a diagonal matrix 
containing the excited mode masses that arise from the KK reduction
 in the
absence of the Higgs interactions. We will refer to this basis
as the ``KK basis''.
  Moreover, unlike in the SM,
since the $Y_u$ does not commute with $M_u$, the
Yukawa matrix for the KK tower is not diagonal in the mass
eigenbasis.

Defining $Y_d,M_d,m_D$ analogously,  we
perform bi-unitary transformations that diagonalize the mass
matrices: \be
L_u M_u R_u^\dagger &=& \mu_u,
\label{eq:mu}\\
L_d M_d R_d^\dagger &=& \mu_d , \ee where $\mu_{u,d}$ are the
(real and diagonal) physically observable fermion mass matrices. The Yukawas
in the mass eigenbasis can be slightly simplified as follows. It
will be useful here and throughout to define the projection matrix
onto zero modes: \be P_0 &\equiv& \bea{cc} 1 & 0 \\ 0 & 0 \ena \en
It is clear that $vY_{u,d} = P_0 M_{u,d}$. The transformation of
this equation into the mass eigenbasis takes the form \be
vL_u Y_u R_u^\dagger &=& L_u \bea{cc}1 & 0 \\ 0 & 0 \ena L_u^\dagger \mu_u \nn\\
vL_d Y_d R_d^\dagger &=& L_d \bea{cc}1 & 0 \\ 0 & 0 \ena L_d^\dagger
\mu_d \label{eq:yukME} \en Notice the presence of only $L$'s on the
RHS. This has the important consequence, mentioned in the introduction,
that flavor-changing in the $W$ and Higgs interactions is controlled
by a single CKM matrix
\be
V_{\rm CKM} &=& L_u P_0 L_d^\dagger \\
V_{uu} &=& L_u P_0 L_u^\dagger = V_{\rm CKM} V_{\rm CKM}^\dagger \\
V_{dd} &=& L_d P_0 L_d^\dagger = V_{\rm CKM}^\dagger V_{\rm CKM}
\label{eq:VCKMdef} \en Equation \ref{eq:yukME} can therefore be
rewritten as \be v Y_{u,d} &=& V_{uu,dd} \mu_{u,d}
\label{eq:yukME2} \en for interactions with neutral Higgses.  We
will find it convenient to work in a gauge where the longitudinal
modes of $W^\pm$ are kept explicitly as the goldstone modes
$h^\pm$.  In the mass eigenbasis, the $h^\pm$ interactions satisfy
$v Y_d = V_{\rm CKM} \mu_d, v Y_u = V_{\rm CKM}^\dagger \mu_u$.

This is very similar to the Standard Model, where all
flavor-changing is controlled by a single CKM matrix.  In
particular, this is a sufficient condition for radiative
corrections to $\bar{\theta}$ and dimension-5 EDM operators to
vanish at 1-loop, as we will discuss in section \ref{sec:edms}. We
note that, unlike in the Standard Model, $V_{\rm CKM}$ is not
unitary.  However, in the limit that the KK masses are large,
mixing between the zero modes and KK modes approaches zero.  In
this limit, $V_{\rm CKM}$ restricted to the KK modes vanishes, and
restricted to the zero modes is exactly the unitary CKM matrix of
the Standard Model.

The structure of the mass matrices has an
interesting and important consequence
for the mixing between zero modes and KK modes.  Compared to models
with doublets localized in the UV, the mixing
between the doublet zero modes and KK modes is greatly enhanced,
while that between singlet zero modes and KK modes is greatly
suppressed.  The reason is that $L_u$ diagonalizes
$M_u M_u^\dagger$ while $R_u$ diagonalizes $M_u^\dagger M_u$, and
these have different orders of magnitude in the off-diagonal components
between zero modes and KK modes.  More precisely, $M_u M_u^\dagger$
 has zero-mode-KK-mode
mixing from the block matrix $v F_u^{(1)} M_u^{(1)}$ while
$M_u^\dagger M_u$ has zero-mode-KK-mode mixing from the block matrix
$(v^2 F_u^{(0)\dagger} F_u^{(1)})$. As we review in section
\ref{sec:notwist}, the size of excited
mode wavefunctions in RS in is  $F_u^{(1)} \sim \sqrt{2}$. Thus, the mixing
for the doublets is therefore of the order $(L_u)_{ij} \sim \sqrt{2}
v/m_{\rm KKf}$, while that for the singlets is of the order
$(R_u)_{ij} \sim \sqrt{2} (v m_0 /m_{\rm KKf}^2)$, where $m_0$ is
the zero-mode mass\footnote{
We have
introduced the notation $m_{\rm KKf}$ to indicate the mass of KK fermions
and $m_{\rm KKg}$ for the gauge bosons.}.

\section{CP and Flavor Constraints}
\label{sec:flavcp}

\subsection{Contributions to $\Phi$ interactions  from Higher-dimensional
Operators}

The vanishing of $\bar{\theta}$ at tree-level depended
crucially on the form of the fermion couplings to $\Phi$.
For this reason, one might worry that higher-dimensional
operators could destroy this result.  This is
not the case, however.  The only higher-dimensional
interaction $\Phi$ can have with a fermion bilinear 
other than $\bar{\Psi} \Psi$ is through
\be
{\cal L} \supset F(\Phi)_{ij} D_M  \bar{\Psi}_i
 \Gamma^M \Psi_j
 \ee
This does not contribute to the $\Psi$ wavefunction for the
following reason.  Since $\Phi$ vanishes on the boundary, all boundary
terms involving $\Phi$ vanish as well. Under the assumption that
$\langle \Phi \rangle$ and the bulk masses (the only sources of
flavor symmetry breaking in our model) are generated spontaneously,
then $D_M \bar{\Psi}_i \Gamma^M \Psi_j$ is the divergence of a
current that is conserved up to a chiral anomaly on the boundary
\cite{ArkaniHamed:2001is}. Since the $\Phi$ field vanishes on the
boundary, and $D_M J^M_{ij}$ vanishes off the boundary, these higher
dimensional operators do not contribute to the fermion wavefunctions
\footnote{Strictly speaking, the brane Yukawas explicitly break the
$U(3)^3$ flavor symmetry corresponding to this current down to the
diagonal subgroup.  However, this breaking occurs on the TeV brane,
sequestered from the $\Phi$ field, and is therefore not a problem.}.

In addition, one might worry that $\Phi$ could couple directly to
$G\tilde{G}$ through higher dimensional operators of the form \be {\cal L}
\supset G_{AB} G_{CD} D_E F(\Phi) \epsilon^{ABCDE}, \ee where
$F(\Phi)$ is a function of $\Phi$ and the form of this operator is
again constrained by the 5D AdS isometries. However, restricting to
the gluon zero modes, which have flat wavefunctions, this operator
is a total derivative of $\Phi$ and therefore integrates to a
vanishing boundary term in the low-energy theory.

Finally, as we have already mentioned in the introduction,
 sequestering $\Phi$ from the TeV brane eliminates the
higher-dimensional operator $\Phi_{ij} \bar{Q}_{Li} H U_{Rj}$.

\subsection{The EDM Operator and $\bar{\theta}$ at One Loop}
\label{sec:edms}

As we saw in section \ref{sec:vertices}, extra dimensional models
generate Hermitian bulk fermion wavefunctions which force
$\bar{\theta}$ to vanish at tree level. In this section, we show
that by also taking the quark doublets to be brane-localized, we
can eliminate $\bar{\theta}$ (and any EDMs) at one loop level.

To begin, let us consider the higher dimension operator
\be {\cal O}_{EDM} &=&   \frac{\bar{d}_L \gamma^{\mu\nu}
\tilde{F}_{\mu\nu} d_R}{\Lambda_{EDM}}, \en which is
phenomenologically relevant for the neutron EDM, and is
constrained by experiment to be $d_n < 10^{-24} e \; {\rm cm}$
\cite{nircp}.

In the past this EDM operator has been quite constraining for
models embedded in warped backgrounds.  For example, in
\cite{agasheperez} the authors assume anarchic brane Yukawas along
with bulk doublet and singlet fermions, and as a result the EDM
operator is generated at one loop via a virtual Higgs.
Consequently, the KK gauge boson masses are constrained to be
$\gtrsim 10$ TeV for that model. As we will now show, with
 the quark doublets on the IR brane, this one loop
contribution vanishes.

Even without evaluating the one loop diagram, we can
immediately see that this contribution explicitly vanishes because its
corresponding spurion vanishes.  The spurion for the down-type
quark EDMs is fixed by flavor symmetries to be \be d_i &\propto&
{\rm Im} \left( L_d (Y_u (M_u^\dagger M_u)^n M_u^\dagger + Y_d
(M_d^\dagger M_d)^n M_d^\dagger) Y_d R_d^\dagger \right)_{ii},
\qquad n  = 0, 1, \dots \label{eq:EDMspur}\ee where the $ii$
subscript labels the down-type quarks (zero and excited modes).
Also, here $n$ denotes some number of mass-squared insertions.

This contribution vanishes for essentially the same reason
that it vanishes in the Standard Model. Using equations
\ref{eq:mu}-\ref{eq:yukME2}, we see that the above term is
\be
d_i &\propto& {\rm Im} \left( V_{\rm CKM}^\dagger \mu_u^{2n+2}
  V_{\rm CKM} \mu_d
  + V_{dd} \mu_d^{2n+2} V_{dd} \mu_d \right)_{ii} \qquad n=0,1,\dots
\en
 It follows from the fact that $V_{ij} V_{ji}^\dagger =
   |V_{ij}|^2$ is real for any matrix $V$ (no sum on $i,j$
implied) that the above term is also real.

Actually, the vanishing of EDMs at one loop also implies the
vanishing of $\bar{\theta}$ at one loop.  This is important because
in most models of spontaneous CP breaking, even if $\bar{\theta}$
vanishes at tree level, one-loop radiative corrections force the
scale of CP breaking up to near the GUT scale.  To see that IR
brane-localized fermions remedy this, recall that
\cite{ellisgaillard} \be \bar{\theta}
&=& \arg \det (M_{u} + \delta M_u) \det (M_{d} + \delta M_d) \nn\\
  &\approx& \im \tr\left(
    \delta M_u / M_{u} +  \delta M_d / M_{d} \right)
    + \ldots, \ee where the $\delta$'s denote contributions from one loop corrections, and we have assumed that $\arg \det(M_u M_d)$ vanishes because
    of hermitian bulk fermion wavefunctions.
the EDM operator, the contribution to $\bar{\theta}$ from, say the
down-type quarks, is proportional to a spurion:
\be \bar{\theta} &=& {\rm Im} {\rm Tr} \left( (Y_u (M_u^\dagger
M_u)^n M_u^\dagger +
   Y_d (M_d^\dagger M_d)^n M_d^\dagger) Y_d M_d^{-1}\right),
\qquad
n  = 0, 1, \dots \\
&=& \sum_i \frac{d_i}{(\mu_d)_i} = 0.  \ee

Contributions to the EDMs and $\bar{\theta}$ from one loop
diagrams with KK $U(1)_Y$ bosons vanish via similar arguments.
Moreover, loops of $SU(2)$ weak gauge bosons do not contribute
because they do not couple to right-handed fields.  Again, this is
only the case because the left-handed doublets are IR-localized.
If this were not the case, then weak bosons would couple to their
right-handed bulk partners and contribute at one loop.

\subsection{$\bar{\theta}$ at Two Loops from $W$'s and Higgses}
\label{sec:TwoLoopThetaBar}

Thus, we have shown that the leading contributions to the EDMs and
$\bar{\theta}$ enter at two loops.  At this loop level, an
evaluation of $\bar{\theta}$ becomes far more complicated because
of the proliferation of diagrams, and so we find it useful to
first analyze the contributions in terms of their
spurions.

We will address diagrams with KK gluon and hypercharge boson ($B$)
contributions in the next
subsection.  Such contributions are sufficiently small only in the
small $t_0$ limit where the KK fermion wavefunctions are diagonal.
We first analyze the contributions from diagrams
with $W^\pm, W^3$, and Higgses, and zero mode gluons and $B$'s. 
In this case, the vertices
are all controlled by a single generalized CKM matrix
$V_{\rm CKM}$, and their contributions to $\bar{\theta}$ are
sufficiently small even for $t_0 \sim 1/100$.

To begin, we note that any diagram renormalizing $M_{u,d}$
necessarily ends with a right-handed fermion line following an
interaction with a Higgs or a neutral gauge boson
(e.g. gluon or  $U(1)_Y$ KK mode, $B^{(1)}$).
However, the zero mode gluon and $B$ interactions
are flavor-respecting so we can  ignore their contributions.  
Also, we note that the right-most interaction cannot be with the
 $W^\pm ,W^3$ or their KK excitations because these fields do not couple to any
right-handed fields.

Thus the dominant loop  contributions to $\bar{\theta}$ will be
from virtual $W$ and $h$ exchange, and the final vertex
insertion should be a Yukawa.
We have already seen in section \ref{sec:yukinteraction}
that all W and Yukawa contributions are formed
out of quark masses and the (generalized) CKM matrices: \be V_{\rm
CKM} &=& L_u P_0 L_d^\dagger \\
V_{uu} &=& V_{\rm CKM} V_{\rm CKM}^\dagger \\ V_{dd} &=& V_{\rm
CKM}^\dagger V_{\rm CKM}.\ee
For example, in mass
eigenstate, the Yukawa interaction between two up-type quarks and a
Higgs field can be written as $L_u Y_u
R_u^\dagger = L_u P_0 L_u^\dagger L_u  M_u R_u^\dagger/v = V_{uu}
\mu_u/v$.
 Note that written in terms of $V$ and $\mu$, each
Higgs propagator is necessarily accompanied by  two factors of $1/v$.

It is important to notice that the right-handed transformation
matrices $R_{u,d}$ do not explicitly appear anywhere, since in
mass eigenstate basis they are all absorbed into mass insertions.  This
implies that the contributions to $\bar{\theta}$ can be written
completely in terms of the fermion masses, the Higgs vev $v$, the
generalized CKM's $V_{uu},V_{dd}, V_{\rm CKM}$, and propagators.
No other terms appear.
We emphasize the similarity of the above
simplification to that of the Standard Model, where all CP and
flavor violation appears through the CKM matrix and quark masses.
Nonetheless, there are two essential differences that complicate
the analysis: 1) there are heavy KK fermions above the $W$ and
$h$ masses so additional factors of their masses do not suppress
the overall contribution,
 and 2) the CKM matrix is not unitary because the doublet zero modes
 mix with left-handed KK singlets.  Thus, $\bar{\theta}$ renormalization occurs at two-loops instead of
three-loops as in the SM.

\begin{figure}[th]
\includegraphics[width=\textwidth]{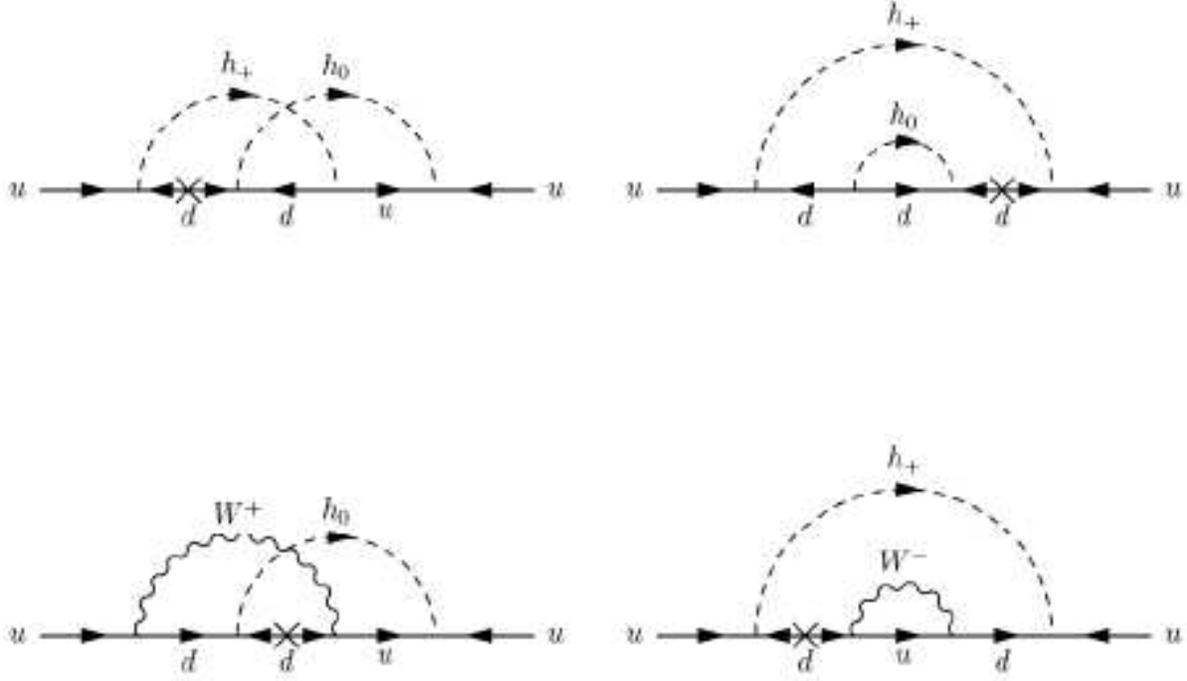}
\caption{There are four topologies of two-loop diagrams that give the
dominant contribution to $\bar{\theta}$, shown above.  The largest of
these corrections come from the crossed-neutral-higgs-charged-higgs
(upper left).  }
\label{fig:thetabardiagrams}
\end{figure}

  There are four distinct topologies of two-loop diagrams with $W$ and/or Higgs loops that
contribute to $\bar{\theta}$, as shown in figure
\ref{fig:thetabardiagrams}: two Higgs lines, either nested or
crossed, and a Higgs line and a W line, either nested or crossed.
   The actual size of the largest
contributions depends on how close the excited mode wavefunctions are
to being proportional to the identity matrix, and therefore on the position
$t_0$ of the CPV field.

Contributions to $\bar{\theta}$ are IR convergent, and dominated
by momenta $\sim m_{\rm KKf}$.
It is
straightforward though somewhat laborious to calculate numerically
the size of contributions to $\bar{\theta}$ from $W,h$ loops.
The largest contribution  comes from the crossed higgs diagram
with a charged Higgs and a neutral Higgs, and a helicity flip on
the first fermion line.
This contribution is given schematically by the two loop integral:
\footnote{The following expression has only {\it three}
interactions because we have canceled away the factor of
$Y_{u,d}M_{u,d}^{-1}$ in the $\bar{\theta}$ expression} \be \delta
\bar{\theta} &=& v^{-4} \tr \int (d^4 p)^2 V_{f_1 f_2}
\frac{m_{f_2}^{2}}{p^2 - m_{f_2}^2} V_{f_2 f_3}
\frac{(p^\mu \gamma_\mu - m_{f_3})
m_{f_3}^2}{p^2 - m_{f_3}^2} V_{f_3 f_1}
\frac{p^\mu\gamma_\mu - m_{f_1}}{p^2 -
m_{f_1}^2} \left(\frac{1}{p^2 - m_{h}^2}\right)^2,
 \nn\\
\label{eq:spurionintegral} \en
 where the $f_i$ run over the zero and excited mode fermions.  Since $f_i$ run over up-like and down-like quarks,
 $V$ denotes the appropriate generalized CKM matrix for its given indices.  For
 example, $V_{f_1 f_2}$ where $f_1$ and $f_2$ are both up-like
 denotes the $f_1 f_2$ component of $V_{uu}$, etc.
Here we have chosen to completely ignore the momentum structure of
the fermion propagators, choosing instead to emphasize the
parametric dependence of the integral.  Also, note the two factors
$m_{f_2}$ in the first propagator, one of which arises from the mass
insertion, and the other which arises from the rewriting a Yukawa
vertex as a CKM multiplied by a mass.  Both factors of $m_{f_3}$
come from this rewriting of neighboring Yukawa interactions. As in
any diagram with two Higgs loops, there is a factor of $1/v^4$. All
diagrammatic contributions will go parametrically as products of
masses and generalized CKMs.

The largest contribution from equation (\ref{eq:spurionintegral})
comes from when 
$f_{1,2}$ are down-like and $f_3$ is up-like (otherwise
one requires three different generations to get a phase
and therefore more mixing angles).  Thus, the three
$V$'s in the diagram are $V_{ud}$, $V_{dd}$, and $V_{du}$.
If any of the masses in the numerator come from zero modes, then
this contribution is immediately very suppressed.  As a result,
since the contribution is maximal, then $f_{2,3}$ must be KK
modes. Consider first the possibility that the remaining fermion
line is also a KK mode.
Then, the entire diagram is suppressed by (at least) \be \delta
\bar{\theta} &\lesssim& (m_{\rm KKf}/v)^4 \; {\rm Im}
\sum_{f_{1,2,3} \in {\rm KKf}} \frac{1}{(4\pi)^4} (V_{ud})_{f_1
f_2} (V_{dd})_{f_2 f_3} (V_{du})_{f_3 f_1} \label{eq:vprod}, \en
where the fermions are summed only over KK modes. As it turns out,
this on its own is enough to be below experimental bounds.
When $f_i,f_j$ are KK modes, $V_{f_i,f_j} \sim (v/m_{KKf})^2$,
as follows from the fact that $V = L P_0 L^\dagger$ and
$L$ contains zero-mode-KK-mode mixing of order
$v/m_{\rm KKf}$.
This suppression and the loop factor just by themselves  already
result in a suppression of
 $(1/4\pi)^4
\left( m_{\rm KKf} / v \right)^4 \left( v^2 / m_{\rm KKf}^2
\right)^3 \sim 3 \times 10^{-7}$. There is additional suppression
from the fact that $f_{1,2,3}$ must all be different, or else the
product of $V$'s has no imaginary piece. This results in a
suppression from mixing between generations. 
 We compute this suppression
numerically.  It is
 due to the fact that the KK mode wavefunctions
$F_{u,d}^{(1)}$ become flavor-symmetric as $t_0 \rightarrow 0$.
We find $V_{f_1  f_2} / V_{f_1 f_1}$ decreases to zero 
approximately as $t_0^{1/2}$ (or faster in some cases)
and thus one requires only $t_0 \lesssim 10^{-4}$ in order
to put the above corrections to $\bar{\theta}$ safely
below experimental constraints.  Finally, there is an
additional suppression beyond what is contained in equation
(\ref{eq:vprod}), due to a partial cancellation from the
interchange of two KK modes. That is, $(V_{ud})_{f_1 f_2}
(V_{dd})_{f_2 f_3} (V_{du})_{f_3 f_1} + f_1 \leftrightarrow f_2$
is real.  If the KK mode masses were exactly equal, this
cancellation would be exact; as it is, it implies only a
cancellation of about $|(m_{f1} - m_{f2})/m_{f1}| \sim 1/10$.

Similar but slightly different considerations imply that the contribution
with a single zero mode internal quark is roughly the same size.
That is, say $f_1 = (u,c,t)$.
As before,
the dominant contribution must have two different KK modes or the
phases cancel.  There is
also a quite drastic suppression from the fact that the zero mode masses
are much smaller than the typical momenta, so the zero mode propagators
are approximately flavor respecting.  Summing over them therefore
approximately removes phases. In particular, this means that the contribution
from a single zero mode internal quark is approximately equal to the
contribution from all KK mode quarks, since the sum
$\sum_j (V_{du})_{KK_1,j} (V_{ud})_{j,KK_2} (V_{dd})_{KK_2,KK_1})
= (V_{du} V_{ud})_{KK_1,KK_2} (V_{dd})_{KK_2,KK_1}$ is real.
Putting in the weak $m$-dependence in the fermion propagators then implies
that these two contributions are roughly equal in magnitude rather than
exactly equal in magnitude (and opposite in sign).
So the contributions with zero mode
internal quarks are small as well. We therefore find that the 
size of radiative corrections to $\bar{\theta}$ from $W$ and higgs
loops is
\be
\delta \bar{\theta} &\sim& 3 \times 10^{-8} t_0
\en
We  note that this result applies when the $SU(2)_L$ gauge group
is confined to the IR brane.  When $SU(2)_L$ is in the bulk, there
are $W$  KK gauge bosons that couple more strongly
to the TeV brane by a factor of $\sqrt{2 k r_c \pi }$
than the zero mode KK gauge bosons do.  Hence the coupling
is enhanced by a factor of $g_w \sqrt{2 k r_c \pi}$ over the
Yukawa interactions, and the radiative corrections are approximately
\be
\delta \bar{\theta} &\sim& 2 \times 10^{-6} t_0
\en
which is below the bound for $t_0 \lesssim 10^{-4}$.  In the
next section, we will consider bigger contributions
to $\bar{\theta}$.  
 The arguments in this section relied upon the fact
that all CPV was controlled in the Higgs and $W$ interactions
by a single $V_{\rm CKM}$ matrix. However, the KK gluons introduce
new sources of CPV that are not controlled by a single
$V_{\rm CKM}$ matrix.    In the next section,
we will provide a different analysis of the two-loop corrections
to $\bar{\theta}$ that will be completely general and applies
to all two-loop radiative corrections, gluons as well as
$W$ and Higgs, but which is valid only in the small $t_0$ limit.

\subsection{$\bar{\theta}$ from Flavor Spurions}

In this subsection we present the estimate  of the
contributions to $\bar{\theta}$ in the small $t_0$ limit
from two loop diagrams containing neutral currents.
The estimate is based on a
flavor spurion analysis we describe in detail in appendix
\ref{app:details}. Our method is also  applicable to
diagrams containing charged currents.

 In the small $t_0$ limit, the full mass matrices take the form
\be
M_{u,d} &=& \bea{cc}
   F_{u,d}^{(0)} & \sqrt{2} {\bf 1} \\
   { \bf 0 } & M_{u,d}^{(1)} \ena
\en
All of the flavor-mixing effects manifestly arise
from the $F_{u,d}^{(0)}$ matrices.  For convenience, we
define
 $f_{u,d}^{(0)} = L_{u,d}^{(0)} F_{u,d}^{(0)} R_{u,d}^{(0)\dagger} $ as the matrix
of eigenvalues of $F_{u,d}^{(0)}$, proportional to the zero mode masses.
It is straightforward to compute numerically that the left-handed diagonalization
matrices $L_{u,d}$ for the full mass matrix then take the form
\be
L_{u,d} &=& \bea{cc} 
   L_{u,d}^{(0)} & -  L_{u,d}^{(0)} \frac{v}{m_{\rm KKf}}\\
   \frac{v}{m_{\rm KKf}} & {\bf 1} \ena
\label{eq:Lapprox}
\en
plus small deviations.
The lower left-hand block of the above matrix
 is only diagonal and not proportional to the identity matrix
because there are small ( ${\cal O}(10 \%)$ ) splittings in the KK fermion
masses.  
We find numerically that given the above structure
for $M_{u,d}$ and the Standard Model values for the zero mode
masses, deviations from equation \ref{eq:Lapprox}  are at 
most ${\cal O}(10^{-13})$ in $L_u$ and even smaller in $L_d$. 
Naively, the contributions to $\bar{\theta}$ from $L_{u,d}$ of the
form \ref{eq:Lapprox} are much greater than $10^{-10}$.
However, as we show in appendix \ref{app:details}, our final
result for the size of $\bar{\theta}$ in the small $t_0$ limit
assuming $L_{u,d}$ takes exactly the form \ref{eq:Lapprox} will be
very small (${\cal O}(10^{-15})$). In fact it is exactly the $O(10^{-13})$ 
deviations in the up-type 
matrices that are responsible for the largest contributions to $\bar{\theta}$. 
We find that at sufficiently small $t_0$, the contributions to
$\bar{\theta}$ are independent of $t_0$ and of the size
\be
\bar{\theta} &\approx& 10^{-13} \qquad \textrm{(small } t_0)
\en

\subsection{Unitarity of $V_{\rm CKM}$}

Due to mixing between zero-modes and KK modes, the full CKM matrix
restricted to zero modes is not exactly unitary.   In models with
UV-localized doublets, this effect is one to two orders of magnitude
below the bound of current experimental constraints
\cite{huber}.    However, because of the larger mixing between zero-modes
and KK-modes
when doublets are on the TeV brane, this effect is much larger
in our model.
The strongest constraints on unitarity can be parameterized by
$\Delta V_i \equiv 1 - \sum_{j=1}^3 |V_{ij}|^2$.
The mixing is approximately ${\cal O}(\sqrt{2} v / m_{\rm KKf})$ for
the doublets, so
\be
\Delta V_i &\lesssim & \frac{2 v^2}{(m^{(1)}_{\rm KKf})^2}\frac{\pi^2}{6}
\en
where the additional factor $\sum_n \frac{1}{n^2} = \pi^2/6$
approximates the contribution from the tower of KK modes
with masses $\sim m_{\rm KKf}^{(n)} \sim n m^{(1)}_{\rm KKf}$.
This is at the level of current experimental constraints
when $m_{\rm KKf} = 9-12 {\rm TeV}$.  This is comparable
to though weaker than electroweak constraints on the model.

\subsection{Flavor-Changing Neutral Currents}

Strong constraints on physics beyond the Standard Model
arise from dimension-6 flavor-changing operators:

\be
{\cal L}_{\Delta F = 2} &\supset&
   \frac{z_{sd}}{\Lambda^2} (\bar{d}_L \gamma_\mu s_L)^2
  + \frac{z_{bd}}{\Lambda^2} (\bar{d}_L \gamma_\mu b_L)^2
\en
There are stronger constraints on the R-L operators
such as $(s_L d_R)(d_L s_R)$. 
However,  because of the small
mixing in the right-handed fermions, this is very
suppressed and the dominant constraints come from
the above operators (see equation \ref{eq:cdrME}, which
is nearly flavor-diagonal when restricted to zero modes).
With $\Lambda \equiv {\rm TeV}$, the constraints on the $z$
coefficients are \cite{YosefNir} \be {\rm Im}(z_{sd}) &\lesssim&
6\times 10^{-9} \qquad
   \epsilon_K \nn\\
z_{bd} &\lesssim& 6 \times 10^{-6} \qquad \Delta m_B 
\en

These operators get contributions from KK gluon exchange.
The KK gluons, like all KK gauge bosons, are peaked at the TeV brane,
with $\chi^{(i)}(1) = 4.72 \approx \sqrt{2 k r_c}$.
In the case of anarchic Yukawas
with the left-handed fields localized on the TeV brane,
the constraints from FCNCs would be much more severe.
 However,
the situation is quite different with sequestered flavor and CP violation.
When flavor-violation is sequestered, the quark-KK-gluon vertices become
approximately flavor conserving in the mass eigenbasis.  We have
already worked out the form of the KK gluon interactions with the
left-handed fermions in the mass eigenbasis in equation
\ref{eq:cdlME}.  Here, we are concerned with interactions with two
zero mode fermions, which are described by the upper-left block 
of \ref{eq:cdlME}: $L_d^{(0)} \left( \chi(1) {\bf 1} 
  + \frac{v}{m_{\rm KK}} D_{dL} \frac{v}{m_{\rm KK}} \right) L_d^{(0)\dagger}$.
The matrix $L_d^{(0)}$ is not exactly unitary, so $L_d^{(0)} L_d^{(0)\dagger}$
has some off-diagonal components that contribute to $z_{sd}, z_{bd}$.
We find numerically that, given our values for the $\nu$'s and
sampling over random $\Phi$ vevs, that 
$(L_d^{(0)} L_d^{(0)\dagger})_{sd} \approx 4 V_{sd} \frac{\delta m}{m_{\rm KK}}
\frac{v^2}{m_{\rm KK}^2}$, where $\delta m \approx 0.1 m_{\rm KK}$ is the
small splitting in the KK fermion mass eigenvalues.  Mixing between the
first and third generation is even more suppressed:
$(L_d^{(0)} L_d^{(0)\dagger})_{bd} \approx 0.1 V_{bd} \frac{\delta m}{m_{\rm KK}}
\frac{v^2}{m_{\rm KK}^2}$.

  Gluon exchange
involves two of the above interactions, so the contribution to
$(\bar{b}_L \gamma_\mu d)^2/\Lambda^2$ in the small $t_0$ limit is approximately
\footnote{ We assume 
 $m_{\rm KKg} \approx \frac{2.45}{\pi} m_{\rm KKf}$, which typically follows from the
equations of motion for KK gauge bosons and KK fermions in RS models.}
\be
z_{bd} &\approx& (1.4)
  (2 k \pi r_c)  g_s^2 \frac{ \left[ \lambda_C^3 0.1 (\delta m/m_{\rm KKf})
 (v/m_{\rm KKf})^2 \right]}
  {m_{KKg}^2} \nn\\
  &\approx& 6 \times 10^{-6}
   \left( \frac{400 {\rm GeV} }{m_{\rm KKg}} \right)^6 
\en
where the 1.4 has been included to account for contributions
from higher KK modes.

The constraints from $\epsilon_K$ are significantly more severe:
\be {\rm Im}(z_{sd}) &\approx&
 (1.4)
  (2 k \pi r_c)  g_s^2 \frac{ \left[ \lambda_C 4 (\delta m/m_{\rm KKf})
 (v/m_{\rm KKf})^2 \right]}
  {m_{KKg}^2} \nn\\
  &\approx& 9 \times 10^{-6}
   \left( \frac{10 {\rm TeV} }{m_{KKg}} \right)^6 
\en

\section{Conclusion}
\label{sec:othermodels}


We have considered a new solution to the strong CP problem
based on spontaneous CP violation in RS models. We sequester
the source of CP violation, which we assume results from the
vev of a scalar field localized near the UV brane, from the
Yukawa interactions in the IR with the Higgs, which we assume
respect a large flavor symmetry.  In order to suppress radiative
contributions to $\bar{\theta}$, we localize the doublet quarks
on the IR brane, which pushes the scale of KK gauge bosons
up to be $\gtrsim$ 15-18 TeV.  We find $\bar{\theta}$ suppressed
to as little as $10^{-13}$.

Let us contrast our model with other models
of spontaneous CPV, beginning with the 4D dual of our warped scenario.
Interestingly, the mechanism that enforces a vanishing strong CP
phase at tree level in the gravitational theory is slightly
different from in the CFT dual. In 5D the hermiticity of bulk
fermion mass matrices is enforced by higher dimensional spacetime
symmetries.  Where the dual is conformal (that is, excluding
energies near CFT breaking), these masses correspond to anomalous
dimensions, which must be hermitian simply because they are
renormalizations of the kinetic terms.

Let us consider the structure of the CFT as we flow down from high
energies. In the UV, CP is a valid symmetry and the theory is
conformal.  At this scale, the UV localized SM fermions are present
in vector-like pairs and as fundamental objects. At the scale
$\Lambda_{CPV}$ the CFT strongly couples, generating a bound state
scalar CP-on $\Phi$ which spontaneously breaks CP.  At this point conformal
symmetry is maximally violated by strong dynamics, so we would
naturally assume that CP phases enter the theory quite generically.
However, as we can compute in the gravity theory, this is not the
case, and strong dynamics  introduces only (hermitian) wavefunction
renormalizations of the (now partly composite) fermions.   Below
this scale, the residual CFT gauge symmetry flows down to the IR
brane, where it strongly couples again, yielding a composite Higgs
scalar and fermion composites corresponding to the heavier SM
fermions. In general, the 4D dual of our model works by keeping all
phases inside wavefunction renormalizations.


This is similar  to what we will call theories of
``wavefunction renormalization.''  
These models
utilize spontaneous CP violation, but subject to symmetries that
effectively sequester CP phases in wavefunction renormalizations
(or higher-dimensional wavefunctions)
alone. Because any renormalization of the kinetic energy term
necessarily implies that the anomalous dimensions are hermitian,
these models forbid strong CP phases at tree level.

For example, in Hiller-Schmaltz \cite{hillerschmaltz}, the supersymmetric
non-renormalization theorem is used to forbid CP phases from seeping
into $\bar{\theta}$, which is only in the superpotential. Because CP
violation occurs above the SUSY breaking scale, phases enter only
into the Kahler potential, and are thus hermitian up to
SUSY-breaking effects. However, in order to obtain appropriately
large weak CP phases, these wavefunction renormalizations must be of
order unity, and so this theory is strongly coupled. As a result,  precise
model-dependent observables are difficult to obtain, and moreover
the inclusion of higher dimension operators generated by these
strong dynamics force the CP violating scale up to around $10^5$
GeV, far above future collider reach.  Because our model is
gravitational, it is completely weakly coupled. Otherwise, the 4D CFT
dual of our theory has many similarities to Hiller-Schmaltz, except
that conformal symmetry effectively replaces supersymmetry in terms
of suppressing phases in the potential.

Our model also has advantages over alternative models of spontaneous
CP violation.  For example, consider the archetypal model of
spontaneous CP violation, proposed by Nelson and Barr
\cite{nelson,nelsonbarr}. This model assumes an exact CP symmetry in
the UV, as well as additional scalars and vector-like fermions at a
high scale charged under the SM flavor group.  After a scalar gets a
CP violating vev, $\bar{\theta}$ is forbidden at tree-level due to
conditions placed on the mass matrix by the GUT representations of
additional fermions.

{}From a model-building point of view, our proposal can be seen as
more minimal than Nelson-Barr. First of all, while the inclusion of
vector-like fermions charged under added flavor symmetries is not
necessary in a 4D model, it follows from the KK reduction in 5D.  In
particular, 5D spacetime symmetries necessarily imply that parity is
a symmetry of the bulk, forcing heavy modes to appear in vector-like
pairs. Furthermore, the mass of the new vector-like states
can be much lower.
Finally, we note that since our model is in RS, we can explain
large hierarchies of mass scales through the warped geometry.




\section*{Acknowledgments}
We would like to acknowledge helpful conversations with
Patrick Meade,
Matthew Schwartz, Patrick Fox, and Nima Arkani-Hamed, and
Gilad Perez for conversations and comments on the draft. 
ALF would like to thank Matthew Baumgart for several
clarifying discussions of electroweak precision constraints.
LR is supported in part by NSF grants
PHY-0201124 and PHY-0556111.  ALF is supported by an NSF Graduate
Research Fellowship.  CC is supported in part by
DOE grant DE-FG02-91ER40654.  Any opinions, findings, and conclusions
or recommendations expressed in this material are those of the authors
and do not necessarily reflect the views of the National Science
Foundation.

\appendix

\section{Bulk Fermion Wavefunctions}
\label{sec:kkmodewf}

\subsection{Wavefunctions Without Twisting}
\label{sec:notwist}

In this subsection we derive bulk fermion wavefunctions in a simply
warped model with no flavor twisting. A right-handed fermion in RS
with bulk mass $m = \nu k$ satisfies the wave equation \be (\d_t^2 +
x^2 - \frac{\nu (\nu + 1)}{t^2} ) f_R &=&0 \en
  The solutions are
\be f_R(t)&=& \sqrt{t} \left( \beta J_{\half + \nu} (x t)
  - \alpha J_{-\half - \nu}(x t) \right)
\en The three conditions that completely determine $\alpha, \beta$,
 and $x$ are the two boundary conditions, $f_R(\epsilon),f_R(1) =0$,
and the normalization $\int_\epsilon^1 f^2_R(t) dt = 1$. It is
useful to work out approximate solutions first. The $t=\epsilon$
condition is, to lowest-order in $\epsilon$, \be 0&=& -
\frac{2^{\half + \nu} x^{-\half - \nu} \alpha \ep^{-\nu}} {
\Gamma(\half - \nu)}
  + \frac{2^{-\half - \nu } x^{\half +\nu} \beta \ep^{1+\nu}}
{\Gamma(\frac{3}{2} + \nu)} \en so $\alpha \ll \beta$ if $\nu >
-\half$, and $\alpha \gg \beta$ if $\nu < -\half$.  The boundary
condition at $t=1$ is therefore \be
0 &\approx& \left\{ \begin{array}{cc} J_{-\half -\nu}(x), & \nu< -\half \\
J_{\half + \nu}(x), & \nu > -\half \end{array} \right.
\label{eq:bcroots}\en The $n$-th root of $J_\mu$ is
well-approximated by $\left( \frac{\mu}{2} - \frac{1}{4} + n \right)
\pi$ in the range $ 5 > \mu > -1/2$.  Over the range $-1 < x< 1$,
this approximates $x$ to within 5 \%.  So, we can take \be x^{(n)}
&\approx& \left\{ \begin{array}{cc}
\left( \frac{\nu}{2} + n \right) \pi, & \nu > -\half \\
\left( -\frac{\nu}{2} - \half + n\right)\pi, & \nu < -\half
\end{array}
\right. \en It is convenient to take $\mu = | \nu + \half|$.
Continuing to take $\ep \ll 1$, the normalization condition for $\nu
> -\half$ is \be 1 &=& \int_\ep^1 f_R^2(t) dt
 = \frac{\beta^2}{2} J_{\mu +1}^2 (x)
\label{eq:normeq} \en
Thus, we have approximately \be f_R(t) &=& \left\{ \begin{array}{cc}
  \frac{ \sqrt{2t} J_{\half + \nu}(x t)}
{ J_{-\half + \nu}(x) }, & \nu > -\half \nn\\
   \frac{ \sqrt{2t} J_{-\half - \nu}(x t)}
{ J_{\half - \nu}(x) }, & \nu < -\half
\end{array} \right.
\en where $x$ satisfy equation \ref{eq:bcroots}. In terms of the
coefficients $\alpha,\beta$, this is \be \nu > -\half &:& \alpha =0,
\qquad \beta =
      \frac{\sqrt{2}}{J_{|\nu - \half|}(x)} \nn\\
\nu < -\half &:& \alpha =
               \frac{\sqrt{2}}{J_{|\nu - \half|}(x)},
       \qquad \beta = 0
\label{eq:coeffsoln} \en

Thus, the left-handed excited mode wavefunctions are given by \be
f_L(t) &=& \left\{
\begin{array}{cc}
  \frac{ \sqrt{2t} J_{-\half + \nu}(x t)}
{ J_{-\half + \nu}(x) }, & \nu > -\half \nn\\
  - \frac{ \sqrt{2t} J_{\half - \nu}(x t)}
{ J_{\half - \nu}(x) }, & \nu < -\half
\end{array} \right.
\en and the zero mode is: \be f_L^0 &=& \sqrt{\frac{1 + 2\nu}{1 -
\ep^{1+2\nu}}} t^{\nu} .\en

\subsection{Wavefunctions With Twisting}

Solving for the wavefunctions can be complicated by twisting.
However, because of the delta function form of $\Phi$, the bulk
fermion wavefunctions can be straightforwardly solved for on to the
left and right of $\phi_0$, and then matched at the junction.

We now show this explicitly.  The following discussion will hold
equally well for $U$ and $D$, so we will drop all $u,d$ subscripts
and let $X = g r \langle \Phi \rangle$.
Rewriting in terms of the $t$ variable, we find that
$\delta(\phi-\phi_0) = krt \delta(t-t_0)$, and so from
Eq.~(\ref{eq:nudef}) we can write $\mu = {\rm
diag}\left(\nu_i\right) + X t_0 \delta(t-t_0)$. where $\nu_i$
ultimately sets the masses of the bulk fermions.
Eq.~(\ref{eq:fullwaveequation}) depicts the first order wave
equations for a bulk fermion written as a matrix differential
operator acting on a matrix wavefunction. For computational
simplicity, let us consider a single column of this equation, which
constitutes a matrix differential operator acting on a vector
wavefunction:
\be \left( \pm \partial_t -
\frac{\mu}{t}\right)
f_{L,R} &=& - x  f_{R,L},\\
\Rightarrow \left(-\partial_t^2 + \frac{\mu(\mu\mp 1)\pm t \mu'}{t^2}\right)
f_{L,R} &=& x^2 f_{L,R}, \label{eq:bulkeqvec}  \ee
where $f_{L,R}$ are 3-vectors, ' denotes differentiation with
respect to $t$,  and $x$ is a mass eigenvalue which
will be determined by the orbifold boundary conditions on the
wavefunctions.

The second order equation in Eq.~(\ref{eq:bulkeqvec}) has a
well-known solution in terms of Bessel functions,
\be f_{L} &=& J_{+-}(t)   a_L +J_{-+}(t)   a_R, \\
f_{R} &=& J_{++}(t)   a_R - J_{--}(t)  a_L, \\
J_{\pm \pm}(t) &=& \sqrt{t}\; {\rm diag}( J_{\pm \frac{1}{2}\pm\nu_i}(x
t)),\ee  where the $a_{L,R}$ are constant 3 vectors defined
piece-wise by \be a_{L,R} &=& \left\{ \begin{array}{ll} a_{L,R}^+ & t> t_0 \\
  a_{L,R}^- & t<t_0 \end{array} \right. \ee
As written, there are precisely four vectors
($a_L^+$,$a_R^+$,$a_L^-$ and $a_R^-$) of initial conditions to
fix. Since one of these is an overall normalization, this leaves
three vector unknowns.

Now if we take $f_R$ to be odd under the orbifold symmetry, then
it vanishes at the orbifold fixed points,
\be f_R(\epsilon)=f_R(1) = 0.
\label{eq:dirichletbc}\ee
These two boundary conditions, along with the zeroth and first
derivative boundary conditions at $t_0$ constitute \emph{four}
boundary conditions that will act to (over-)constrain the
\emph{three} initial conditions for the fermion wavefunctions.
Since this over-constrains the system, this puts a constraint on
$x$ and fixes the allowed masses to a discretum.  Enforcing equation
\ref{eq:dirichletbc} allows us to eliminate the $a_L$'s and write
\be
f_R(t) &=& \left\{ \begin{array}{ll} \left(\frac{J_{--}(1)J_{++}(t)-J_{++}(1)J_{--}(t)}
    {J_{--}(1)}\right)  a_R^+   & t>t_0 \\
   \left( \frac{J_{++}(t)J_{--}(\epsilon)- J_{--}(t)J_{++}(\epsilon)}
   {J_{--}(\epsilon)} \right)  a_R^- & t<t_0 \end{array} \right.
\label{eq:frsimp}
\en

Next, let determine the matching conditions at the third brane.
Because the first order equation in Eq.~(\ref{eq:bulkeqvec})
relates the first derivative of the fermion wavefunction to a
delta function, we know that there is jump discontinuity at $t_0$.
Moreover, without loss of generality, we can regulate this jump
with a linear interpolating function, so \be f_{L,R}(t_0) =
\frac{1}{2}(f_{L,R}(t_0^+)+f_{L,R}(t_0^-)), \ee where $\pm$ denote
limiting values from the right and left.  Our solutions will of
course be insensitive to this regulation.

The first boundary condition is then easily obtained by
integrating the first order wave equation over a small
neighborhood around $t_0$:
\be -f_R |_{t_0^-}^{t_0^+} - \int_{t_0^-}^{t_0^+}  \frac{\mu}{t}
f_R dt &=& - x \int_{t_0^-}^{t_0^+} f_R dt.\ee
Since the integrand on the RHS is not singular at $t_0$ except for
a discontinuity and it is integrated over an infinitesimal range,
its contribution vanishes. In fact only the delta function
component of $\mu$ survives the integration, yielding
\be f_R(t_0^+)-f_R(t_0^-) &=& -\frac{1}{2}X(f_R(t_0^+)+f_R(t_0^-)),\\
\Rightarrow f_R(t_0^+) &=& R f_R(t_0^-),\\
R &=& \frac{1 - X/2} {1 + X/2 },\\
&\approx& 1 - X \en where the last relation holds in the limit
that $X$ is small.

Next, we find the second boundary condition at $t_0$ by
integrating the second order wave equation over an infinitesimal
neighborhood around $t_0$:
\be -f_R' |_{t_0^-}^{t_0^+} + \int_{t_0^-}^{t_0^+}
\frac{\mu(\mu\mp 1)\pm t \mu'}{t^2} f_R dt &=& x^2
\int_{t_0^-}^{t_0^+} f_R dt. \ee
Plugging in for $\nu$ we find that
\be f_R'|_{t_0^-}^{t_0^+} \pm X f_R'(t_0) -\frac{1}{t_0}\left(\{
{\rm diag}(\nu_i),X \} + X t_0 \delta(0)\right)f_R(t_0) &=&0.
 \ee
Here we have introduced a $\delta(0)$ which can be easily removed
by plugging in the first order wave equation evaluated at $t_0$:
\be -f_R'(t_0)-\frac{1}{t_0}\left({\rm diag}(\nu_i) + X t_0
\delta(0) \right) f_R(t_0) &=& -x f_L(t_0). \ee
Combining these equations and plugging in for $f_L(t_0)$ in terms
of $f_R(t_0^+)$ and $f_R(t_0^-)$ from the first order wave
equation, we finally obtain the two boundary conditions for the
fermion wavefunctions at $t_0$:
\be f_R(t_0^+) &=& R f_R(t_0^-),\\
 f'_{R}(t_0^+)
&=& R^{-1} f'_{R}(t_0^-)
   + S  f_{R}(t_0^-), \\
R &=& \frac{1 - X/2} {1 + X/2 },\\
 S &=& \frac{1}{ 1 - X/2 }
  \frac{ \{{\rm diag}(\nu_i) , X \}}{t_0}
 \frac{1}{ 1 + X/2 }.
\en

The above equations, along with the two orbifold boundary
conditions constitute four linear matrix equations which can be
combined (after plugging in $f_{L,R}$) in order to yield an equation
relating $a_R^+$ and $a_R^-$
\be
a_R^+ &=& \frac{J_{--}(1)}{J(1)} R 
\frac{J(\epsilon)}{J_{--}(\epsilon)} a_R^-
\label{eq:arplusarminus}
\en
and a single
condition on $a_R^+$:
 \be G(x) a_R^+ &=& 0, 
\ee 
where we have defined the matrix functions
\be J(t) &\equiv& J_{--}(t)
J_{++}(t_0) - J_{++}(t) J_{--}(t_0), \\
G(x)&\equiv& (-\partial_{t_0}J(1)  R   J(\epsilon)+ J(1) R^{-1}
 \partial_{t_0}J(\epsilon) + J(1) S
J(\epsilon)) J_{--}^{-1}(\epsilon). \ee The matrix
$G(x)$ does not have zeros for any arbitrary $x$, and so this
linear equation forces $x$ to take on a discretum of values. It is
straightforward to numerically obtain the zeros of $\det G(x)$,
thus yielding the fermion mass eigenvalues $x_i$. Then, the
$a_R^+$ corresponding to the mass eigenvalue $x_i$ is the null 
eigenvector of $G(x_i)$, and $a_R^-$ is given by equation
\ref{eq:arplusarminus}. 
Thus from equation \ref{eq:frsimp} we have the fermion bulk wave functions.

\section{Analysis of Flavor Spurions in small $t_0$ limit}
\label{app:details}

 To begin, we go to mass
eigenbasis, which is advantageous because this separates out the
flavor-mixing, which occurs only at interaction vertices, from
other sources of flavor-breaking. We consider the (broken) chiral
flavor symmetry $G_F = (SU(3)_L \times SU(3)_R)^{N_{KK}}$ which
acts separately on each KK level, treating the symmetry-breaking terms
as spurions charged under this symmetry. Also, let us first
consider diagrams with only neutral current interactions, and generalize
our argument later to include both charged and neutral currents. Thus
for the present, 
loop diagrams necessarily contain only up-like or only down-like
quarks, and we can first consider only with the part of
$G_F$ which acts on the down-like quarks.  We will consider a
general element of $G_F$, $Z_L \times Z_R \times K_L \times
K_R$, where $Z_L$ ($Z_R$) is a  special unitary transformation that acts
on the left-handed (right-handed) zero modes and $K_L$ ($K_R$) is a 
special unitary
transformation that act on the first excited KK left-handed (right-handed) 
modes.   Since
the mass matrix, $L_d,R_d$, $C_{dR}$, and $C_{dL}$ all 
break this symmetry, we treat their zero mode and KK blocks as
spurions.  We will now describe these interactions in some detail
in the small $t_0$ limit and catalogue the spurions under the 
broken flavor symmetry $G_F$.

To begin, consider the interactions between fermions and gluons
before KK reducing: \be {\cal L} &\supset& g_3 \sqrt{\pi} A_\mu \left(
\bar{q}_i \gamma^\mu q_i \delta(t-1)+
   \bar{U}_i \gamma^\mu U_i + \bar{D}_i \gamma^\mu D_i \right)
\en Let us denote the wave-function for the $(n)$-th KK gauge
boson by $\chi_A^{(n)}(t)$, in a convention where \be A_\mu(x,t)
&=& \sum_{n=0}^N A_\mu^{(n)} \frac{\chi_A^{(n)}(t)}{\sqrt{r_c}}
\en
In this convention, the zero mode wavefunctions $\chi_A^{(0)}(t) =
1/\sqrt{\pi}$ are flat, the excited mode wavefunctions satisfy
$\chi_A^{(n)}(1) \approx \sqrt{2 k r_c}$ where $n>0$. Because the
zero mode wavefunction is a constant and the fermion KK modes are
orthonormal, the interactions with the zero mode gauge boson are
exactly flavor-respecting.  This is of course the result of the
exact gauge symmetry.

As it turns out, the  KK gluon interactions also possess an approximate flavor
symmetry. To see this, consider the coupling of the $(n)$-th KK
gluon to the zero mode and first KK mode fermions.  We will write
this interaction in terms of the $u_{L,R}$ basis defined in
equations \ref{eq:nicebasis}-\ref{eq:mudef}, where the approximate $U(1)^3$ 
flavor symmetry is respected by
the KK mode masses.  We refer to this basis as the ``KK basis'',
to distinguish it from the mass eigenbasis where the full mass matrix including
mixing with the zero modes and KK modes is diagonal.  

In the KK basis, the gluon coupling to
the $u_R$ is given by the integral of the KK wavefunctions over
the fifth dimension:
\be C_R^{(n)} &=& g_s \sqrt{\pi}\int \chi_A^{(n)}(t)\bea{cc}
  F_u^{(0)}(t)^\dagger F_u^{(0)}(t)   &  F_u^{(0)}(t)^\dagger F_u^{(1)}(t) \\
 F_u^{(1)}(t)^\dagger F_u^{(0)}(t)  & F_u^{(1)}(t)^\dagger F_u^{(1)}(t)
 \ena dt.
 \ee
Here each $F$ is a three by three flavor matrix.  For the case of
$u_L$, the gluon interaction matrix is simpler since the doublet
fields have no KK excitations: \be C_L^{(n)} &=& g_s \sqrt{\pi}
\int \chi_A^{(n)}(t)\bea{cc}
 \delta(t-1) &  0 \\
 0  & F_{u'}^{(1)}(t)^\dagger F_{u'}^{(1)}(t)
 \ena dt,
 \ee
 where $F_{u'}$ denotes the bulk wavefunction corresponding to the
 left handed KK excitation of the right handed singlet
 $u_R^{(0)}$.  

The coupling of the fermions to a KK gluon can be 
 understood qualitatively by
decomposing the gluon wavefunction into a flat
``UV'' piece that is constant and an oscillating ``IR'' piece that
vanishes in the UV:
 \be
\chi_A^{(n)}(t) &=& \chi_{\rm UV}(t) + \chi_{\rm IR}^{(n)}(t) \\
\chi_{\rm UV}(t) &=& -\frac{1}{\sqrt{8 k r_c}} \\
\chi_{\rm IR}^{(n)}(t) &\approx& \sqrt{2 k r_c t} \sin\left( t \frac{2n-1}{2} \pi\right)
\label{eq:kkgwfapprox}
\ee
The ``UV'' piece is flat and therefore, by orthonormality of the
fermion wavefunctions, flavor-universal.    
For
$C_R^{(n)}$, the IR contribution roughly yields products of
$F_u$'s evaluated near the IR brane, and thus has the same order
of magnitude as the Yukawa matrices.  
We thus obtain qualitatively 
\be C_R^{(n)} &\approx& g_s \sqrt{\pi}\left( -\frac{1}{\sqrt{8kr}} {\bf
1} + \sqrt{2kr} Y_u^\dagger Y_u\right). 
\label{eq:cr} \ee 
This implies that the KK gluon vertices that mix right-handed 
zero modes fermions 
and right-handed 
KK mode fermions are suppressed by the zero mode masses, which we
will find later to
be essential to making radiative corrections to $\bar{\theta}$ small.

The left-handed partners of the KK singlets have wavefunctions
that vanish on the IR brane, and their coupling to the KK 
gluon does not receive the full $\sqrt{2k r_c}$ volume factor:
\be C_L^{(n)} &\approx& g_s \sqrt{\pi}\bea{cc}
  \sqrt{2 kr} {\bf 1}    &  0 \\
 0  & {\cal O}(2)
 \ena.
\label{eq:cl}
 \ee

For radiative corrections to $\bar{\theta}$ to be small, we 
need small $t_0$.  We can describe the interactions approximated in equations
\ref{eq:cr} and \ref{eq:cl} more precisely in 
 this small $t_0$ limit, where the  KK mode fermion wavefunctions
are independent of $\Phi$ but the zero mode fermion wavefunctions 
are not.  
The essential point will be that the only sources of flavor mixing
are the left-handed diagonalization matrices $L_{u,d}^{(0)}$ 
\emph{for the zero modes} and that the only new flavor-violating 
spurions beyond those in the Standard Model that can contract with $L_{u,d}^{(0)}$ 
are suppressed by $(v/m_{KKf})^2$ or the usual mass matrix for the zero modes.  
Thus, we now wish to demonstrate that all flavor-mixing in the gluon
vertices can be pushed in the mass eigenbasis into $L_u^{(0)}$ and $L_d^{(0)}$.

First, we show that $L_{u,d}^{(0)}$ are also the only sources
of flavor mixing in the gluon interaction vertices. In particular,
$R_{u,d}^{(0)}$ can be absorbed into the diagonalizations of the
zero mode masses (and equivalently $F_{u,d}^{(0)}$) so
that they manifestly do not contribute new sources of flavor-mixing.  We will
restrict our attention to the gluon interaction with down quarks for the
moment, to avoid writing extra indices.  
First, note that since the KK fermion wavefunctions are 
diagonal in the present limit, their interactions with the KK gauge
boson are also diagonal in the KK basis.  We can denote this
by defining $D_{dR} = C_{dR,11}  $, so $D_{dR}$ is a diagonal
matrix. Second, consider the interactions of the KK gauge
bosons with a zero mode fermion and a KK mode fermion.  These are given by
$ C_{dR,01} = g_s \sqrt{\pi} \int_\epsilon^1 \chi(t) F_d^{(0)\dagger}(t)
  F_d^{(1)}(t) $.  However, the integral receives only a
negligible contribution from
$t<t_0$, and above that $F_d^{(0)\dagger}(t) = F_d^{(0)\dagger}(t_0)$
times a diagonal matrix, according to equation \ref{eq:zeromodeint}. 
This implies that at any point $t>t_0$, $F_d^{(0)\dagger}(t)$ is related to 
$F_d^{(0)\dagger}(1)$ by a diagonal matrix.
Thus, $C_{dR,01} = F_d^{(0)\dagger}(1) D_{dR}'$,
where we have defined  another diagonal matrix $D_{dR}'$.
Finally, note that because the
zero modes are UV localized and do not oscillate in the IR,
they see the IR piece $\chi_{\rm IR}$ of the KK gluon wavefunction as
essentially a spike near the IR brane, and thus the qualitative approximation
of $C_{dR}$ we gave earlier is quantitatively very good for the zero modes:
$C_{dR,00} \approx g_s \sqrt{\pi} (-\frac{1}{\sqrt{8 k r_c}} {\bf 1} + 
\sqrt{2 k r_c}  F_d^{(0)\dagger}F_d^{(0)})$. Furthermore, the effect of
deviations from this approximation are suppressed by the zero
mode wavefunctions.  The exact interaction contains
the flavor-violating contribution
$F_d^{(0)\dagger } D_{dR;0} F_d^{(0)}$, again with
 $D_{dR;0}$ a diagonal matrix.
In the mass eigenbasis, this is
$f_d^{(0)} L_d^{(0)} D_{dR;0} L_d^{(0)\dagger} f_d^{(0)}$.
The small off-diagonal pieces of $ L_d^{(0)} D_{dR;0} L_d^{(0)\dagger}$
are then further suppressed by the hierarchy in zero mode
masses, making them completely negligible.
Thus, in the KK basis, $C_{dR}$ takes the form
\be
C_{dR} &=& 
g_s \sqrt{\pi}   \bea{cc} \chi(\epsilon) {\bf 1} + \chi(1) F_d^{(0)\dagger} 
    D_{dR;0} F_d^{(0)} &
   F_d^{(0)\dagger} D_{dR}' \\
  D_{dR}' F_d^{(0)} & D_{dR} \ena
\en

Consider now the rotation to mass eigenbasis.  The right-handed mixing
between zero modes and KK modes is suppressed by zero mode masses and
$v/m_{\rm KKf}$, where we find numerically the result
\be
R_d &=& \bea{cc} R_d^{(0)} & 
  - R_d^{(0)} F_d^{(0)\dagger} \frac{\sqrt{2}v^2}{m_{\rm KKf}^2}\\ 
 \frac{\sqrt{2}v^2}{m_{\rm KKf}^2} F_d^{(0)}
 & {\bf 1} \ena
\en
We now use this to rotate $C_{dR}$ to the mass eigenbasis:   
\be
C_{dR} &\rightarrow&
   \bea{cc} \chi(\epsilon) {\bf 1} + 
  \chi(1) f_d^{(0)2} & f_d^{(0)} L_d^{(0)} D_{dR}' \\
   D_{dR}' L_d^{(0)\dagger} f_d^{(0)} & D_{dR} \ena 
  + {\cal O}(\frac{v^2f_d^{(0)2}}{m_{KKf}^2})
\label{eq:cdrME}
\en
We have not written out some terms that are higher order
in $\frac{v^2}{m_{\rm KK}^2}$.  
This is because
they can be written as spurions we have already included (times 
some diagonal matrices) and thus can only give contributions that
are suppressed by $v^2/m_{\rm KK}^2$ over those we are including.
The above form  demonstrates  our claim that $R_d^{(0)}$
does not explicitly appear in the gluon interactions, and thus the only source
of flavor-mixing in the down-type interactions is $L_d^{(0)}$.  
The matrix of KK gauge boson interactions with the left-handed fields
is simpler. In the KK basis, it takes the form we have already
derived:
\be
C_{dL} &=& \bea{cc} \chi(1) {\bf 1} & 0 \\
   0 & D_{dL} \ena 
\en
and thus in the mass eigenbasis does not contain any new flavor-breaking spurions
beyond $D_{dL}$ and those in $L_d$.  Explicitly, it transforms to
\be
C_{dL} &\rightarrow& 
   \bea{cc}
    L_d^{(0)} \left( \chi(1) {\bf 1} + \frac{v}{m_{\rm KK}} D_{dL}
   \frac{v}{m_{\rm KK}} \right) L_d^{(0)\dagger}  &
   L_d^{(0)} \frac{v}{m_{\rm KK}} \left( \chi(1) {\bf 1} - D_{dL} \right) \\
    \left( \chi(1) {\bf 1} - D_{dL} \right) \frac{v}{m_{\rm KK}} L_d^{(0)\dagger}&
   \chi(1) \left( \frac{v}{m_{\rm KK}^2} + D_{dL} \right) \ena
\label{eq:cdlME}
\en

In particular, the mass matrix supplies the three
spurions $f_d^{(0)}, \sqrt{2} {\bf 1},$ and $M^{(1)}$, $L_d$
supplies the spurions $L_d^{(0)}$ and $ \frac{v}{m_{\rm KKf}}$,
and $C_{dR}$ and $C_{dL}$ supply the spurions $F_d^{(0)\dagger}
D_{dR}', D_{dR},$ and $D_{dL}$. Under the chiral flavor symmetry, these
spurions transform like \be
f_d^{(0)} &\rightarrow& Z_L F_d^{(0)} Z_R^\dagger \\
\sqrt{2} {\bf 1} &\rightarrow & Z_L \sqrt{2} K_R^\dagger \\
f_d^{(0)} L_d^{(0)} D_{dR}' &\rightarrow&
    Z_R F_d^{(0)\dagger} D_{dR}' K_R^\dagger \\
D_{dL} &\rightarrow & K_L D_{dL} K_L^\dagger \\
M_1 &\rightarrow& K_L M_1 K_R^\dagger \\
L_d^{(0)} &\rightarrow& Z_L L_d^{(0)} Z_L^\dagger \\
\frac{v}{m_{\rm KKf}} {\bf 1}&\rightarrow& Z_L \frac{v}{m_{\rm
KKf}} {\bf 1} K_L^\dagger \en Next, we estimate the value of
$\bar{\theta}$ from a two loop diagram with neutral currents.
First, we note that in order to obtain a contribution to
$\bar{\theta}$, a diagram must necessarily include four factors of
$L_d^{(0)}$. Otherwise the phase in the first flavor-mixing $L_{dij}^{(0)}$ is
precisely canceled by mixing back through the inverse factor
$L_{dji}^{(0)\dagger}$. The flavor symmetry then fixes which
spurions can interleave around these four factors.

The only spurions that transform under $Z_L$ aside from
$L_d^{(0)}$ are $f_d^{(0)}$ and $\frac{v}{m_{\rm KKf}}$.  The
first of these transforms on the right under $Z_R$, so it must be
squared in order to appear between two $L_d^{(0)}$'s. Note that
these spurions are nothing more than the zero mode masses, and it
is exactly this argument that implies such a large suppression of
$\bar{\theta}$ in the Standard Model.

 The second of these, $\frac{v}{m_{\rm KKf}}$,
transforms on the right under $K_L$, so it also must be squared in
order to appear between $L_d^{(0)}$'s. Thus, it also gives a
suppression comparable to $(m_b/v)^2$.

Next, if the external mode is a zero mode, then at least one of
the spurions must be proportional to $f_d^{(0)}$ (to see this
explicitly, note that all of the spurions charged under $Z_R$ are
proportional to $f_d^{(0)}$), which then cancels with the
$m_d^{-1}$ in the expression for corrections to $\bar{\theta}$.
Thus, the minimum suppression allowed by the above argument is
$(v/m_{\rm KKf})^7 = 3 \times 10^{-12}$. Finally, there is the fact that
the overall contribution must be proportional to the Standard
Model phase invariant $\Delta^{(4)} \equiv {\rm Im} \left(
L_{d12}^{(0)} L_{d22}^{(0)\dagger} L_{d21}^{(0)}
L_{d11}^{(0)\dagger} \right) \sim 10^{-4}$ (see section \ref{sec:values}). 
 This puts the final
contributions from two neutral currents (or more specifically, two
gluons) safely below the experimental limits on $\bar{\theta}$. We
note that, when all of the loops are from KK gauge bosons, each
vertex contains a large coupling $g_5 \chi(1) = g_4 \sqrt{2 k \pi
r_c} \lesssim \sqrt{70}$, which is very close to the
non-perturbative regime. In particular, each additional loop is
suppressed only by a factor of $g_4^2 \frac{2 k \pi
r_c}{(4\pi)^2}$.

We can now extend the above argument to loops including charged
currents. In this case, we must consider a separate flavor
symmetry acting on up-type quarks and down-type quarks: $G_F =
G_{Fu} \times G_{Fd} \equiv \left( SU(3)_L \times SU(3)_R
\right)^{N_{KK}}_u \times \left(SU(3)_L \times SU(3)_R
\right)^{N_{KK}}_d $, since charged currents will generate
diagrams with both up-like and down-like quarks.
It is crucial that the only new flavor-mixing spurions from
charged currents arise from $V_{\rm CKM} = L_u L_d^\dagger$, which
in the small $t_0$ limit takes the form \be V_{\rm CKM} &=&
  \bea{cc}
   L_u^{(0)} L_d^{(0)\dagger} & L_u^{(0)} \frac{v}{m_{\rm KKf}} \\
   \frac{v}{m_{\rm KKf}} L_d^{(0)\dagger} &
   \frac{v^2}{m_{\rm KKf}^2} \ena
\en These new spurions therefore transform like: \be V_{\rm
CKM}^{(0)} \equiv L_u^{(0)} L_d^{(0)\dagger}
    &\rightarrow& Z_L^u V_{\rm CKM}^{(0)} Z_L^{d \dagger} \\
L_d^{(0)} \frac{v}{m_{\rm KKf}} &\rightarrow&
  Z_L^d  L_d^{(0)} \frac{v}{m_{\rm KKf}} K_L^{u \dagger} \\
L_u^{(0)} \frac{v}{m_{\rm KKf}} &\rightarrow&
  Z_L^u  L_u^{(0)} \frac{v}{m_{\rm KKf}} K_L^{d \dagger} \\
 \frac{v^2}{m_{\rm KKf}^2} &\rightarrow&
  K_L^u \frac{v^2}{m_{\rm KKf}^2} K_L^{d \dagger}
\en

Thus, in diagrams with charged currents, we have two different
flavor-mixing spurions, $L_u^{(0)}$ and $L_d^{(0)}$. In a totally
general case with two flavor-mixing spurions, we could get
$\bar{\theta}$ contributions with only two insertions of mixing
matrices. However, the present case is far from generic, since
there are
no spurions other than $V_{\rm CKM}^{(0)}$ that are charged under
both $Z_L^u$ and $Z_L^d$. Thus, in order to make an invariant from
both $L_u^{(0)}$ and $L_d^{(0)}$, one must connect them with
additional factors of $L_{u,d}^{(0)}$.  We can therefore apply our
earlier arguments about spurions charged under $Z_{L,R}^{u,d}$.
The phase invariant $\Delta^{(4)}$ constructed from $L_d^{(0)}$ is
approximately the same if we replace $L_d^{(0)}$ by $V_{\rm
CKM}^{(0)}$ since most of the CKM matrix comes from the down-like
quarks of our model. In particular, the largest new invariant that
one can construct, even with charged currents, is still $\left(
\frac{v}{m_{\rm KKf}} \right)^7
   \Delta^{(4)} \approx 10^{-16}$.

There is one last subtlety if the charged currents are Higgs
(equivalently, longitudinal modes of the $W$ zero mode), since its
coupling to KK modes in terms of the above spurions contains a
large factor of $m_{\rm KKf}/v$.  However, there is also no volume
factor since the Higgs is IR localized and has no KK modes, so the
overall largest contribution for instance from a charged Higgs and
a KK gluon is of order $\delta \bar{\theta} \approx
 g_4^2 \left( \frac{2 k \pi r_c}{(4\pi)^4} \right)
  \left( \frac{v}{m_{\rm KKf}} \right)^5 \Delta^{(4)} \lesssim 10^{-15}$.

\section{Precision Electroweak Constraints}
\label{sec:constraints}

Here, for completeness, we review the constraints from
precision electroweak measurements
with a combination of bulk and brane fields.
We will compute all constraints
by matching our UV theory to an effective 4d theory with only Standard Model
fields. These constraints have been discussed previously in the
literature, see e.g.
\cite{csaba,witek,witek2, Burgess:1993vc}.
The most highly constrained operators in the flavor-symmetric
electroweak Lagrangian are
\be
{\cal L}_{EW} &\supset&
   \frac{z_{WB}}{\Lambda^2} (h^\dagger \sigma^a h) W_{\mu\nu}^a B^{\mu\nu} +
   \frac{z_{h}}{\Lambda^2} |h^\dagger D_\mu h|^2 \nn\\
   &&+
   \frac{z^s_{hl}}{\Lambda^2} i (h^\dagger D^\mu h)(\bar{l} \gamma_\mu l) +
   \frac{z^t_{hl}}{\Lambda^2} i (h^\dagger \sigma^a D^\mu h) (\bar{l} \gamma_\mu
   l)\nn\\
   &&+
   \frac{z^s_{hq}}{\Lambda^2} i (h^\dagger D^\mu h)(\bar{q} \gamma_\mu q) +
   \frac{z^t_{hq}}{\Lambda^2} i (h^\dagger \sigma^a D^\mu h) (\bar{q} \gamma_\mu
   q)\nn\\
   &&+
   \frac{z_{hu}}{\Lambda^2} i (h^\dagger D^\mu h)(\bar{u} \gamma_\mu u) +
   \frac{z_{hd}}{\Lambda^2} i (h^\dagger D^\mu h)(\bar{d} \gamma_\mu d) +
   \frac{z_{he}}{\Lambda^2} i (h^\dagger D^\mu h)(\bar{e} \gamma_\mu e)
\label{eq:ewlag}
\en

We present the constraints on the coefficients of these operators
based on the $\chi^2$ comparison to the data performed in \cite{witek}.
We consider only the 9 operators in equation (\ref{eq:ewlag})
\footnote{
The constraints from 4-fermion operators
are negligible for the cases we discuss.
Operators involving only quarks are weakly constrained
and when $SU(2)_L$ is in the bulk operators involving leptons
may be made small by localizing the leptons sufficiently in the UV.
When $SU(2)_L$ is on the brane, the doublet leptons must be IR localized
and thus the operators $(\bar{l}\gamma^\mu l)(\bar{l} \gamma_\mu l)$
and $(\bar{l} \gamma^\mu \sigma^a l)(\bar{l} \gamma_\mu \sigma^a l)$
are larger.  However, in this case there are no KK modes
of $W,Z$ bosons, and the effect of including constraints on these operators
is small.}
the deviation from the $\chi^2$ for the standard model
\be
\Delta \chi^2 &\equiv& \chi^2 - \chi^2_{\rm SM}
\en
where $\chi^2_{\rm SM}$ is the $\chi^2$ value with all dimension-6
operators set to zero. Choosing the 9 operator coefficients to 
minimize $\chi^2$
gives $\chi^2_{\rm min} = \chi^2_{\rm SM} - 12.5$.

\subsection{Contributions from KK modes}

\begin{figure}[th]
\includegraphics[width=\textwidth]{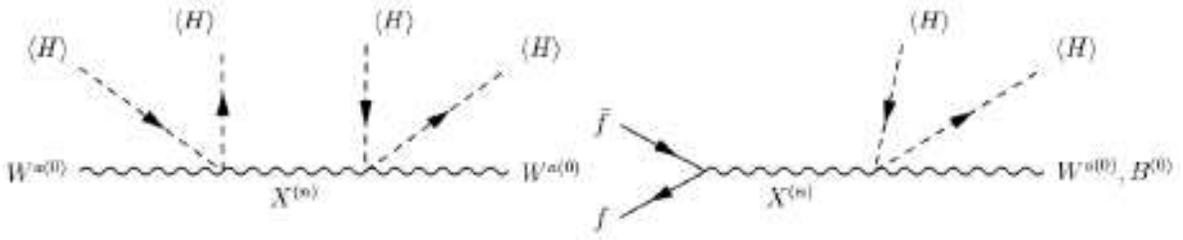}
\caption{Diagrams that give contributions to electroweak observables at tree-level.
The left diagram contributes to $T$ and the right contributes
to fermion-Higgs operators. $X^{(n)}$ indicates KK modes of bulk gauge
fields. }
\label{fig:pewcdiagrams}
\end{figure}

The contribution to fermion-Higgs operators occurs, as with most of
the constrained higher-dimensional operators, at tree-level.
The diagrams that generate $z_{h\psi}$ are shown in
figure \ref{fig:pewcdiagrams} on the right.  The contribution from
hypercharge KK modes is\footnote{See eq 5.2 in \cite{agashesundrum}.}
\be
\frac{z^s_{h\psi}}{\Lambda^2 } &\approx&
  \frac{1}{8} g'^2 z_v^2 Y_\psi \times \left(
\bea{rr} \frac{-k\pi r_c}{2} + \half  &
  \psi \textrm{ IR localized } \\
  \frac{ 1}{4} - \frac{1}{4 k \pi r_c} & \psi
  \textrm{ UV localized} \ena \right) \\
\frac{z^t_{h\psi}}{\Lambda^2} &\approx&
    \frac{1}{4} g^2 z_v^2 \times \left(
\bea{rr} \frac{-k\pi r_c}{2} + \half  &
  \psi \textrm{ IR localized } \\
  \frac{ 1}{4} - \frac{1}{4 k \pi r_c} & \psi
  \textrm{ UV localized} \ena \right)
\en
where $Y_\psi$ is the hypercharge of $\psi$ and
$z_v \equiv \frac{1}{\mu_{\rm TeV}}
 \approx \frac{\sqrt{6}}{m_{KKg}}$ is the
$z$-position of the IR brane.  The contributions to
$z^t_{h\psi}$ vanish when $SU(2)_L$ is on the
IR brane.  Consider first the case where $SU(2)_L$ is
in the bulk.  Then the constraints from the above operators
depend on which fermions reside on the IR brane.

\vspace{2cm}

\begin{figure}[t]
\includegraphics[width=0.47\textwidth]{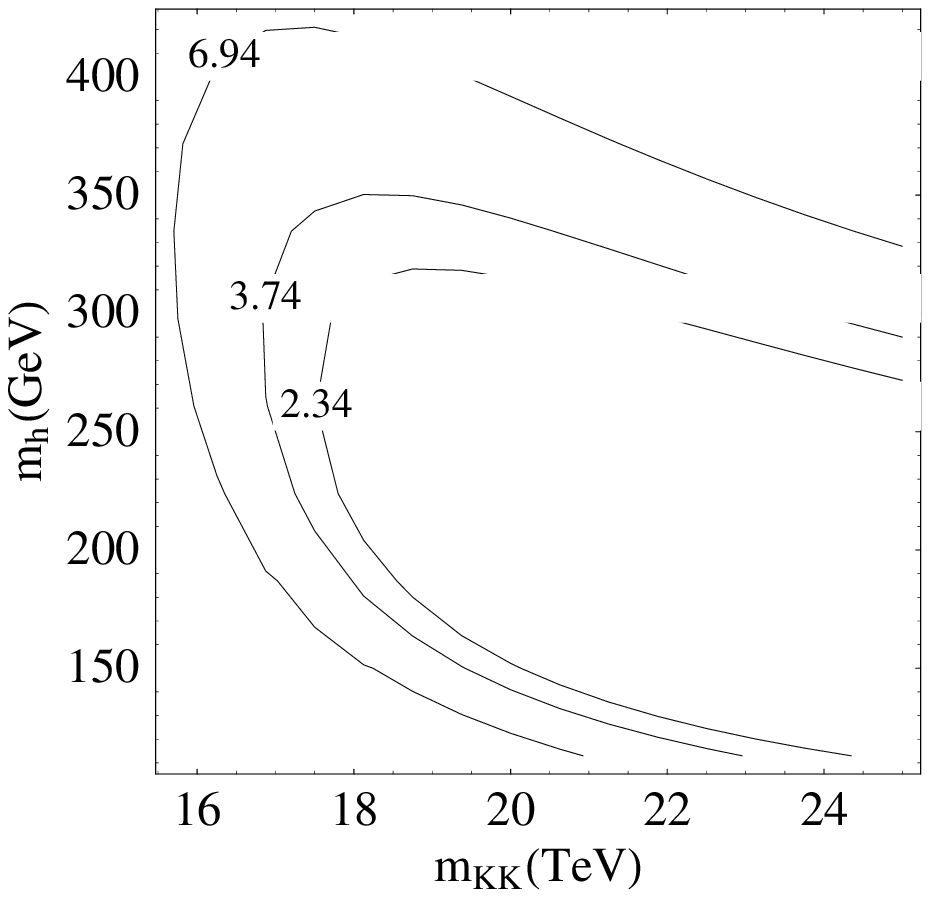}
\includegraphics[width=0.47\textwidth]{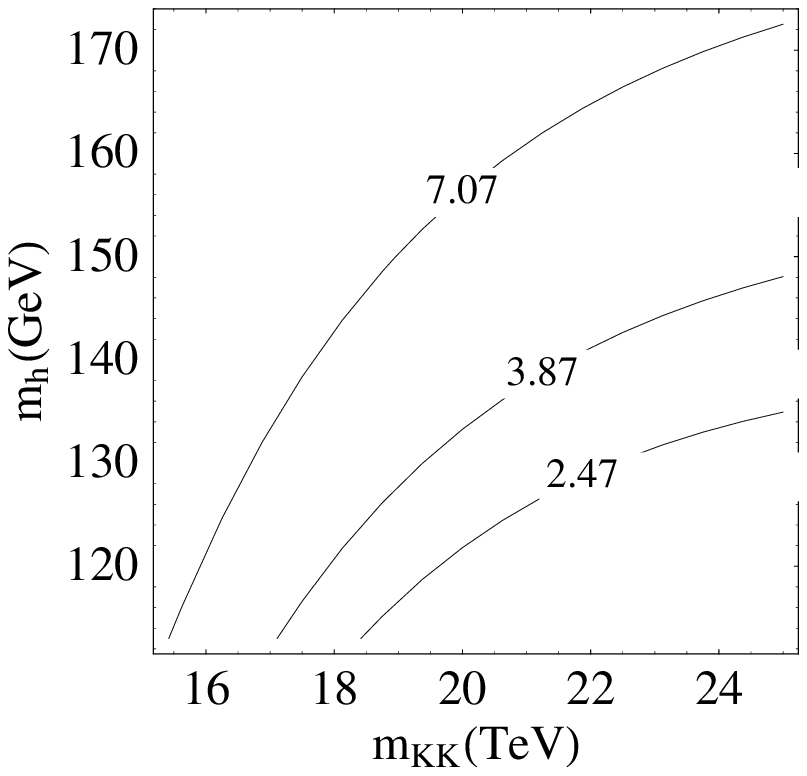}
\caption{Plots of $\Delta \chi^2 = \chi^2 - \chi^2_{\rm SM} $
contours (90\%, 95\%, 99\%) when the doublets are all on the brane.
The left(right) plot shows the case when $SU(2)_L$ is in the
bulk (on the TeV brane).}
\label{ewcontours}
\end{figure}

The leading contribution to $T$ is shown in figure \ref{fig:pewcdiagrams} on
the left, and
have been calculated before \cite{agashesundrum}\footnote{Their equation 4.7.}.
The $U(1)_Y$ KK mode contribution and the $SU(2)_L$ KK mode contribution give
\be
\Pi_{33}(0) &\approx&  -\frac{v^2}{4}\frac{ (k\pi r) (g^2+g'^2) (v z_v)^2}{8}
 \left( 1 - \frac{1}{k\pi r_c}\right) \nn\\
\Pi_{11}(0) &\approx& -\frac{v^2}{4}\frac{ (k\pi r) (g^2) (v z_v)^2}{8}
 \left( 1 - \frac{1}{k\pi r_c}\right)
\en
Recall
\be
T &\approx &\frac{ - 16 \pi (\Pi_{33}(0) - \Pi_{11}(0))}{v^2 e^2}\nn\\
  &\approx& -\frac{16 \pi}
   {v^2 g'^2\cos^2 \theta_W } (\Pi_{33}(0) -\Pi_{11}(0))
\en
It is related to the operator coefficient by
\be
\frac{z_h}{\Lambda^2} &=& -\frac{g'^2 \cos^2\theta_W}{2\pi v^2} T \nn\\
   &=& - \frac{g'^2}{4} k\pi r_c z_v^2
  \left( 1 - \frac{1}{k\pi r_c} \right)
\en
With $SU(2)_R$ in the bulk, the leading order piece in the volume
factor $(k \pi r_c)$ cancels.
With $m_h  = 113$, we find, for $\Delta \chi^2 \equiv
 \chi^2 - \chi^2_{\rm SM}$,

\vspace{1cm}
\begin{center}
\begin{tabular}{|c|c|c||c|c|c|}
\hline
$SU(2)_L$ & $SU(2)_R$  & $L$ doublet & $m_{KKg,\Delta \chi^2 = 4.5}$
  & $m_{KKg,\Delta \chi^2 = 9}$   \\
\hline
In Bulk & In Bulk & In Bulk &  26.0 & 22.7  \\
\hline
In Bulk & In Bulk & On Brane & 26.5 & 23.4  \\
\hline
In Bulk & Not Present & In Bulk &   28.0 & 24.4\\
\hline
In Bulk & Not Present & On Brane &  22.4 & 20.0\\
\hline
On Brane & Not Present & On Brane &   15.6 & 14.2\\
\hline
\end{tabular}
\end{center}
\vspace{1cm}

though as we have mentioned in order to address the strong CP problem
we have $L$ doublets on the brane in our models.
For 9 fit parameters, $\Delta \chi^2 = 
\chi^2 -(\chi^2_{\rm min,9} +12.5) = 4.5 (9.0)$
for 95\% (99\%) confidence. Certain linear combinations of
the fermion-Higgs operators are effectively equivalent, via a field
redefinition \cite{witek2,agashesundrum},
to contributions to $S$ and $T$.  When the fermions
are on the brane, this contribution to $S$ is effectively negative
\cite{csaba}, whereas new physics usually gives a positive contribution
to $S$. Therefore, it is not difficult, though somewhat ad hoc, to
introduce new particles that alleviate these constraints. A less ad
hoc option is to consider a heavy higgs, which increases $S$ at the
price of decreasing $T$.  Contours are shown in figure \ref{ewcontours}
for all doublets on the TeV brane, for $SU(2)_L$ i) on the brane or
ii) in the bulk, without $SU(2)_R$ gauged. Minimizing $\chi^2$
over $m_{KKg}, m_h$ gives $\chi^2_{\rm min,2} =\chi^2_{\rm SM} - 2.13$
in the former case and
$\chi^2_{\rm min,2} =\chi^2_{\rm SM} - 2.25$ in the latter case.
In both cases, electroweak
constraints force $m_{\rm gauge}^{(1)} \gtrsim 15 {\rm TeV}$.

\section{Domain Walls}

Since CP is a discrete symmetry that is spontaneously broken
at the scale $E_{\rm CP} = \mu_{\rm TeV} t_0^{-1}$, domain walls
will form as the universe cools and passes through a phase
transition. It is therefore important that this phase
transition happen only before inflation and not after.
During inflation, the universe rapidly cools to
the de Sitter temperature $T_{\rm dS} = E_{\rm inf}^2/M_{\rm pl}$,
where $E_{\rm inf} = V^{1/4}_{\rm inf}$ is the energy scale of
inflation.  If $E_{\rm CP} > T_{\rm dS}$, then the phase
transition will occur during inflation.  There is also a danger
that the universe will reheat above $E_{\rm CP}$, and then
domain walls would form again after inflation, as the universe
cooled from above to below $E_{\rm CP}$.  In order to avoid this,
it is necessary that the reheating temperature $T_{\rm rh} <
E_{\rm CP}$.  If reheating is efficient, then $T_{\rm rh} =
E_{\rm inf}$, but in general the reheating temperature
can be much lower if the efficiency
is very small, i.e. $T_{\rm rh} = \epsilon_{\rm eff} E_{\rm inf}$.
Both constraints are satisfied for $E_{\rm CP} = 3 {\rm TeV}$
if for example $E_{\rm inf} \approx 10^{11} {\rm GeV}$ and
$\epsilon_{\rm eff} \approx 10^{-8}$.

\section{KK Fermion Mass Matrix}
\label{sec:kkmassmatrix}

The complete mass matrix for the fermions involves the masses
of the kk modes, which get contributions from the yukawa
interactions.  In order for $\bar{\theta}$ to truly vanish
at tree-level, it is important that these mixing terms do not
ruin the reality condition on the mass matrix determinant.
In fact, it is easy to see that they do not, with or without
doublets in the bulk.  Consider first the case with doublets in
the bulk. Then,
being very explicit, before going to the yukawa eigenbasis,
 the mass matrix for the zero modes and the first two
excited modes takes the following form:
$$
\bea{c} u_L^{(0)} \\
   u_L^{(1)} \\
   u_L^{'(1)} \\
   u_L^{(2)}\\
   u_L^{'(2)} \ena^T
\bea{ccccc}
  vF^{q(0)\dag} F^{u(0)} & 0  & vF^{q(0)\dag} F^{u(1)} & 0 & vF^{q(0)\dag} F^{u(2)} \\
  vF^{q(1)\dag} F^{u(0)} & M_Q^{(1)} & vF^{q(1)\dag} F^{u(1)} & 0  & vF^{q(1)\dag} F^{u(2)} \\
          0  & 0 & M_U^{(1)} & 0 & 0\\
  vF^{q(2)\dag} F^{u(0)} & 0 & vF^{q(2)\dag} F^{u(1)} & M_Q^{(2)} & vF^{q(2)\dag} F^{u(2)} \\
  0 & 0 & 0 & 0 & M_U^{(2)} \ena
\bea{c} u_R^{(0)} \\
   u_R^{'(1)} \\
   u_R^{(1)} \\
   u_R^{'(2)} \\
   u_R^{(2)} \ena
$$
where
$M_U^{(i)}$ denotes a diagonal, real mass matrix of the KK masses.
This matrix is hermitian, thanks to the many empty entries,
as long as $M_U^{(i)}, M_Q^{(i)}$, are hermitian and
$F^{q(0)\dag} F^{u(0)}$ has real determinant.  Notice that $F^{q(i)\dag} F^{u(j)}$ does not
affect the determinant of this matrix be hermitian for any $(i,j)$
except for $(0,0)$.

When doublets are on the brane, the mass matrix is even simpler.
$$
{\cal L} \supset \bea{c} u_L^{(0)}  \\
   u_L^{'(1)} \\
   u_L^{'(2)} \ena^T
\bea{ccc}
  vF^{u(0)} & v F^{u(1)} &  F^{u(2)} \\
          0 & M_U^{(1)} & 0\\
  0 & 0 & M_U^{(2)} \ena
\bea{c} u_R^{(0)} \\
   u_R^{(1)} \\
   u_R^{(2)} \ena
$$
which clearly has real determinant.

\newpage

\end{document}